\documentclass[amssymb,prc,floatfix,amsmath,twocolumn,longbibliography]{revtex4-2}
\usepackage{graphicx,bm,float}
\usepackage{bm,color}
\usepackage{epstopdf}
\usepackage{natbib}
\newcommand{\qn}{q_{0_{\cal N}}}
\newcommand{\etal}{{\em et al.}}
\newcommand{\A}[1]{$^{#1}$}
\newcommand{\vlk}{V_{\text{low-}k}}

\newcommand{\epinu}{$(e_{\pi},e_{\nu})$\,}
\newcommand{\enupi}{$(e_{\nu},e_{\pi})$\,}
\newcommand{\eff}{e$^2$fm$^4$\,}
\newcommand{\bet}{$B(E2: 2^+_1\to 0^+_1)$\,}
\newcommand{\bett}{$B(E2: 2_3^+\to 0_2^+)$\,}
\newcommand{\bef}{$B(E2: 4^+_1\to 2^+_1)$\,}
\newcommand{\beq}{$B(E2: 4^+_1\to 2_1^+)/B(E2: 2_1^+\to 0_1^+)$\,}
\newcommand{\betm}{B(E2: 2^+\to 0^+)}
\newcommand{\bl}{\begin{Large}}
\newcommand{\el}{\end{Large}}
\newcommand{\be}{\begin{equation}}  
\newcommand{\ee}{\end{equation}}
\newcommand{\bg}{\begin{gather}}  
\newcommand{\eg}{\end{gather}}
\newcommand{\ba}{\begin{eqnarray*}}
\newcommand{\ea}{\end{eqnarray*}}

\newcommand{\hw}{$\hbar \omega \,$}
\newcommand{\hbw}{\hbar \omega}

\newcommand{\ie}{{\it i.e.,\ }}
\newcommand{\q}{$\langle 2q_{20}\rangle\,$}

\DeclareUnicodeCharacter{2212}{-}
\begin{document}
\title{Quadrupole dominance in the light Sn and in the Cd isotopes }
\author{ A.~P.~Zuker}
\affiliation{Universit\'e de Strasbourg, IPHC, CNRS, UMR7178 Strasbourg, France}

\email{andres.zuker@in2p3.cnrs.fr}
\begin{abstract}
  \begin{description}
  \item[Background] The \bet of the Sn isotopes for $N\le 64$ exhibit
    enhancements hitherto unexplained. The same is true for all the Cd
    isotopes.
\item[Purpose] To describe the electromagnetic properties of the Sn and
  Cd isotopes.
\item[Method]  
    Shell-model calculations are performed with a minimally renormalized realistic
    interaction, supplemented by quasi- and pseudo-SU3 symmetries and
    Nilsson-SU3 self-consistent calculations. Special care is devoted
    to the monopole part of the Hamiltonian.
  \item[Results] (1) Shell-model calculations with the neutron
    effective charge as single free parameter describe well the \bet
    and \bef rates for $N\le 64$ in the Cd and Sn isotopes. The former
    exhibit weak permanent deformation corroborating the prediction of
    a pseudo-SU3 symmetry, which remains of heuristic value in the
    latter, where the pairing force erodes the quadrupole
    dominance. Calculations in $10^7$- and $10^{10}$-dimensional
    spaces exhibit almost identical $B(E2)$ behavior: A vindication of
    the shell model. (2) Nilsson-SU3 calculations describe \bet
    patterns in \A{112-120}Cd and \A{116-118}Sn isotopes having
    sizable quadrupole moment of non-rotational origin denoted as
    q-vibrations.  
  \item[Conclusion] A radical reexamination of traditional
    interpretations in the region has been shown to be necessary, in
    which approximate symmetries involving the quadrupole force and a
    high quality monopole Hamiltonian play a major role. What emerges
    is a bumpy but coherent view.
  \end{description}
\end{abstract}

\date{\today} \maketitle 
\section{Introduction}
\begin{figure}[b]
\begin{center}
\includegraphics[width=0.5\textwidth]{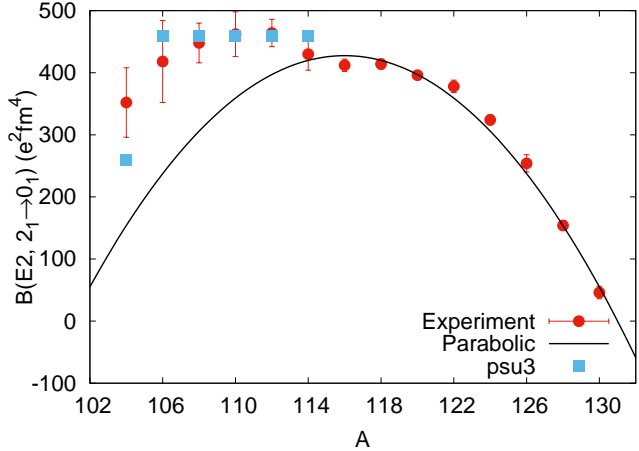}
\end{center}
\caption{\label{fig:SnBE2} The experimental \bet for the Sn isotopes
  from compilations~\cite{be22016}, compared with some arbitrary
  parabolic shape and pseudo-SU3 results to be explained here
  (squares).}
\end{figure}

All nuclear species are equal, but some are more equal than
others. The tin isotopes deserve pride of place, because $Z=50$ is the
most resilient of the magic numbers, because they are very numerous,
and many of them stable, starting at $A=112$. For these, accurate data
have been available for a long time. As seen in Fig.~\ref{fig:SnBE2}, a
parabola accounts very well for their \bet trend, except at
\A{112-114}Sn.
That these early results (Jonsson~\etal~\cite{jonsson1981}) truly
signaled a change of regime became evident through work on the
unstable isomers, starting with the measurement in \A{108}Sn by Banu
\etal ~\cite{banu2005}. A flurry of activity followed~\cite{
  vaman2007sn, cederkall2007sub, ekstrom2008sn, kumar2010enhanced,
  bader2013quadrupole, doornenbal2014intermediate,
  kumar2017noevidence}, from which a new trend emerged in which the
parabola, characteristic of a seniority scheme, gives way to a
platform, predicted by a pseudo-SU3 scheme (the squares).  Here, I am
going a bit fast, to follow the injunction of Montaigne: start by the
last point (``Je veux qu'on comance par le dernier point'' Essais II
10)~\cite[p. 296]{montaigne}. To slow down, I note that the idea to
associate the plateau with pseudo-SU3, originated in a study of the
cadmium isotopes, where quadrupole dominance is stronger and its
consequences more clear-cut. Therefore, it is convenient to study the
Cd and Sn families together. Section~\ref{sec:th} provides the
necessary tools.
\section{\label{sec:th}Theoretical framework}

\begin{figure}[t]
\begin{center}
\includegraphics[width=0.5\textwidth]{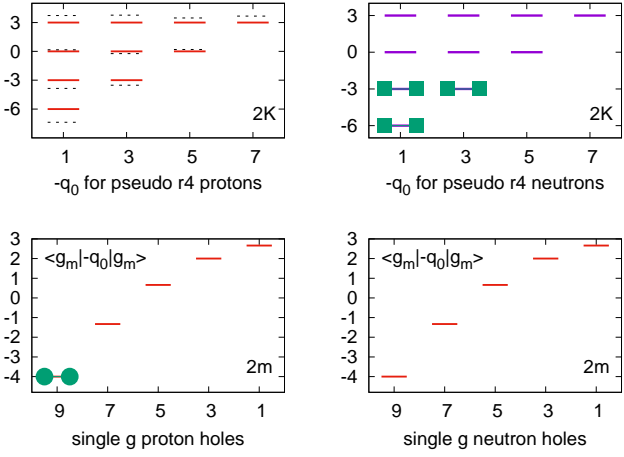}
\end{center}
\caption{\label{fig:SP} The $q_0$ diagrams for SP spaces adapted to the 
Cd and Sn isotopes. The (dimensionless) $q_0=$\q values correspond to 
single particle and hole occupancies for the pseudo $r_4$ (P$r_4)$ and 
$g$ cases respectively. The minus sign is an artifact to make occupancies 
start from the bottom. The figure illustrates the \A{104}Cd configuration: 
circles for holes and squares for particles. The dashed lines are for \q 
from exact diagonalization.}
\end{figure}

The basic idea is inspired by Elliott's SU3
scheme~\cite{elliotta,elliottb} and consists in building intrinsic
determinantal states that maximize $q_0$, the expectation value of the
quadrupole operator $\hat q_0=2q_{20}$, \ie
$q_0$=\q~\cite{Q,rmp,nilssonSU3}.  Fig.~\ref{fig:SP} implements the
idea for \A{104}Cd ($Z=48,\, N=56$).  The single-shell contribution 
(S) of the $g_{9/2}\equiv g$ proton orbit (S$g$) is given
by~Eq.\eqref{q0S} (with changed sign for hole states) with $m=9/2$
being the first unfilled orbit for prolate deformation. For the
neutron orbits, the pseudo-SU3 scheme
~\cite{arimapsu3,hecht1969,nilssonSU3} (P generically, P$r_p$ for
specific cases) amounts to assimilating all the orbits of a major
osillator shell of principal quantum number $p$, except the largest
(the $r_p$ set) into orbits of the $p-1$ major shell. In our case the
$sdg$ shell has $p=4$, and $r_4$ is assimilated into a $pf$ shell. As
the $\hat{q_0}$ operator is diagonal in the oscillator quanta
representation, maximum \q is obtained by orderly filling states
$(n_z\, n_y\, n_x)=(300),(210),(201)\ldots (012),(003)$, with
$q_0=$\q=$2n_z-n_y-n_z=$ 6, 3, 3, 0, 0...-3, -3, as in
Fig.~\ref{fig:SP}. Using $q(n)$ for the cumulated $q_0$ value (e.g. 24
for \A{104}Cd in Fig.~\ref{fig:SP}), the intrinsic quadrupole moment
then follows as a sum of the single-shell (S) and pseudo-SU3 (P)
contributions
\begin{gather}
q_0(S)=2\langle r^2C_{20}\rangle =\sum_m(p+3/2)\frac{j(j+1)-3m^2}{2j(j+1)} \label{q0S}\\
q_0(P)=q(n),~~~Q_0(SP)=[(8e_{\pi}+q(n)e_{\nu})b^2]~{\rm e}{\rm fm}^2\label{Q0SP}
\end{gather}
where I have introduced effective charges and recovered dimensions 
through $Q=b^2q$ with 
$ b^2\approx 41.467/\hbw~ {\rm fm}^2$ and
$\hbw=45A^{-1/3}-25A^{-2/3}\label{hbw}$.  To adapt  Eq.~\eqref{Q0SP}
to Sn, simply drop the S part (\ie the $8e_{\pi}$ term). 

\begin{table}[t]
  \caption{\label{tab:SP} \bet estimates for \A{98+n}Cd in \eff from 
Eq.\eqref{be2}. B20sp uses naive $2q(n)_n$ cumulated pair occupancies 
 diagonalization of $\hat q_0$ in the $pf$ shell \ie strict SU3, 
with $(e_{\nu},e_{\pi})=(1.2,1.5)$.  The B20SP numbers use (full) 
$2q(n)_f$ from diagonalization of $\hat q_0$ in the $r_4$ space, 
$(e_{\nu},e_{\pi})=(1.0,1.4)$. The $b^2$ values range from 4.83 fm$^2$ 
for $A=98$ to 4.99 fm$^2$ for $A=110$. Experimental values (B20e) 
for \A{102-104}Cd are taken from~\cite{Cd100-104} and \cite{Cd102-104}, 
and from compilations~\cite{be22016} for \A{106-110}Cd.}
\begin{ruledtabular}
\begin{tabular}{ccccccc}
$A$&100&102&104&106&108&110\\
$n$&2&4&6&8&10&12\\
$2q(n)_n$&12&18&24&24&24&24\\
$2q(n)_f$&14.8&21.6&29.5&30.0&29.6&29.3\\
B20e&$<$560(4)&562(46))&779(80)&814(24)&838(28)&852(42)\\
B20sp&327 &536&799&808&817&825\\
B20SP&317&511&795&824&817&813\\
\end{tabular}
\end{ruledtabular}
\end{table}

To qualify as a Bohr Mottelson rotor, $Q_0(SP)$ must coincide with the
intrisic spectroscopic $Q_{0s}$ and transition  $Q_{0t}$ quadrupole
moments, defined through (as, e.g, in Ref.~\cite{nilssonSU3})
\begin{gather}
Q_{spec}(J)=<JJ\vert3z^2-r^2\vert JJ> \nonumber \\ 
Q_{0s}=\frac{(J+1)\,(2J+3)}{3K^2-J(J+1)}\,Q_{spec}(J), \quad K\ne1  
\label{bmq}
\end{gather}
\begin{gather}
 B(E2,J\rightarrow J-2)= \frac{5}{16\pi}\,e^2|\langle 
JK20| J-2,K\rangle |^2 \, Q_{0t}^2\label{bme2}\\  
K\ne 1\ \betm=Q_0(SP)^2/50.3 ~{\rm e}^2{\rm fm}^4\label{be2}  
\end{gather}
To speak of deformed nuclei two conditions must be met: \beq=1.43
(the Alaga rule from Eq.~\eqref{bme2}), and the 
``quadrupole quotient'' rule, $Q/q$ which follows from Eqs.\eqref{bmq} and
\eqref{bme2} by equating $Q_{0s}\approx Q_{0t}$ (for $K=0$):
                                                                 
\begin{gather}\label{qQ}
50.27B(E2:2^+\to
0^+)/(3.5Q_{spec})^2=(Q/q)^2\approx 1\end{gather}

Full verification demands calculations but Eq.~\eqref{be2} can be
checked directly by inspecting Fig.~\ref{fig:SP} as done in
Table~\ref{tab:SP}, that will be analyzed once the shell-model
results are in.

These results rely on diagonalizations in spaces defined by
$(g^{X-u}{r}_{4}^u)_{\pi}(g^{10-t}r_4^{n+t})_{\nu}$, $X=8$ for Cd and
10 for Sn. The proton ($u$) and neutron ($t$) excitations are
restricted to have $u+t \leq M$.  The calculations were done for
$utM=000$ (the case in Fig.~\ref{fig:SP}), 111, 101 and 202 using
$\vlk$ variants~\cite{vlk} of the precision interaction
N3LO~\cite{N3LOa} (denoted as I in what follows) with oscillator
parameter \hw= 8.4 MeV and cutoff $\lambda=2$ fm$^{-1}$.  As a first
step the monopole part of I is removed and replaced by single-particle
energies for $^{100}$Sn from Ref.~\cite{gemo} referred to as GEMO for general
monopole: a successful description of particle and hole spectra on
magic nuclei from O to Pb, in particular consistent with the
analysis of Ref.~\cite{Sn100} for \A{100} Sn.  Specifically
$\epsilon_j=$ 0.0, 0.5, 0.8 and 1.6 MeV.  for $j=$ 5/2, 7/2, 1/2, and 3/2,
respectively.)

The I interaction is then renormalized by increasing the
$\lambda \mu=20$ quadrupole and $JT=01$ pairing components by
q$\times$10\% and p$\times$10\%, respectively, and subject to an
overall 1.1 scaling to account for renormalizations of other origins.
The resulting interactions are called I.q.p.  According to
Ref.~\cite{mdz}, the quadrupole renormalization (due to 2\hw
perturbative couplings) amounts to 30\%, a theoretically sound result,
empirically validated by the best phenomenological interactions in the
$sd$ and $pf$ shells.  By the same token the effective charges in 0\hw
spaces are estimated as \enupi= (0.46, 1.31), as confirmed in
Refs.~\cite{BE2FeCr, PhysRevC.69.064304}.  For the pairing component,
perturbation theory is not a good guide, but comparison with the
phenomenological interactions demands a 40\% increase~\cite{mdz,rmp}.
It follows that I.3.4 and \enupi= (0.46, 1.31) should be taken as
standard for full 0\hw spaces.

As I will be working in very truncated bases, which demand large
effective charges, renormalizations should be implemented to account
for polarization mechanisms that involve excitations to the $g$
shell. Proton jumps will contribute to $e_{\nu}$ and are expected to
have greater impact than the corresponding neutron jumps, rapidly
blocked by the $(r_4^{n+t})_{\nu}$ particles. As a consequence, I set
$e_{\pi}=1.4$, a guess close to the standard value, and let $e_{\nu}$
vary, thus becoming the only adjustable parameter in the
calculations, a choice validated later in Sec.~\ref{sec:SM}.

\section{The light C\MakeLowercase{d} isotopes}
\label{sec:cd}
\begin{figure}[t]
  \begin{center}
    \includegraphics[width=0.5\textwidth]{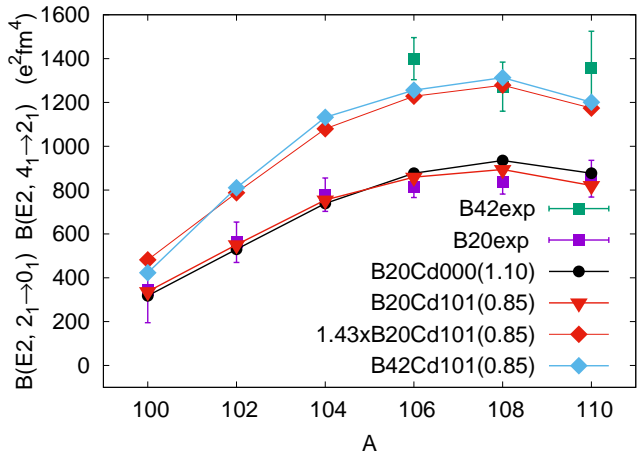}
\end{center}
\caption{\label{fig:Cdbe2} Experimental and calculated $B(E2)$ rates
  for the Cd isotopes, for different $utM$ values. \bet values 
  from~\cite{Cd100-104},\cite{Cd102-104}, and ~\cite{be22016}. \bef
  values from~\cite{nudat2}. In parenthesis $(e_{\nu})$,
  $e_{\pi}=1.40$ is fixed.}
\end{figure}

In Fig.~\ref{fig:Cdbe2} it is seen that $utM=000$ and 101 give the
same results provided $e_{\nu}$ is properly chosen. There is little
difference between $utM=111$ and $utM=101$ because as soon as neutrons
are added they block the corresponding jumps, as mentioned above.

The calculation exhibits near perfect agreement with the Alaga rule:(
\beq$\approx 1.43$\bet). In the figure it is shown for $utM=101$ but
it holds as well for 000 and 111. The more stringent quadrupole
quotient rule Eq.~\eqref{qQ} yields an average $Q/q=0.96$ for
\A{106-110}Cd, corroborating the existence of a deformed region.
As announced immediately after Eq.~\eqref{qQ}, Eq.~\eqref{be2} can be
checked directly by inspecting Fig.~\ref{fig:SP}, as done in
Table~\ref{tab:SP} describing the ``back of an envelope'' SP estimates.   
Note that the naive form of P used so far (in $q(n)_n$ and B20sp) is
supplemented by the more accurate $q(n)_f$ and B20SP using fully
diagonalized values of \q.  The remarkable property of the $r_4^n$
space that produces four identical $q(n)_s$ values for $n=6-12$ has
already been put to good use in Ref.~\cite{Q} and Ref.~\cite[Fig. 38,
TableVII]{rmp}. In the present case, it is seen to do equally well.

It follows that the very simple estimates suggested by
Fig.~\ref{fig:SP} are quantitatively reliable and can be associated with
stable deformation in Cd. In Sn, the same estimates will remain
reliable but they cannot be associated with stable deformation.  A
paradox examined in Sec.~\ref{sec:beqano}.

My interest in the Cd isotopes was stimulated by the work of Schmidt
and collaborators~\cite{schmidt2017}, an extension of
Ref.~\cite{Cdth}, where the $v3sb$ interaction (a renormalized version
of the charge dependent Bonn potential~\cite{CDB}) was defined. The
quadrupole coherence in these studies was clearly identified, but the
\bet plateau was missed. I obtained the interaction from Nadya
Smirnova~\cite{nadya}, and reproduced their results. 
If $v3sb$ is made monopole free and the single particle
spectrum replaced by the  GEMO value for I.3.4, the
\hbox{\bet} pattern becomes identical to that of Fig.~\ref{fig:Cdbe2}. 
\section{The light S\MakeLowercase{n} isotopes}\label{sec:sn}
\begin{figure}[t] 
  \begin{center}
\includegraphics[width=0.5\textwidth]{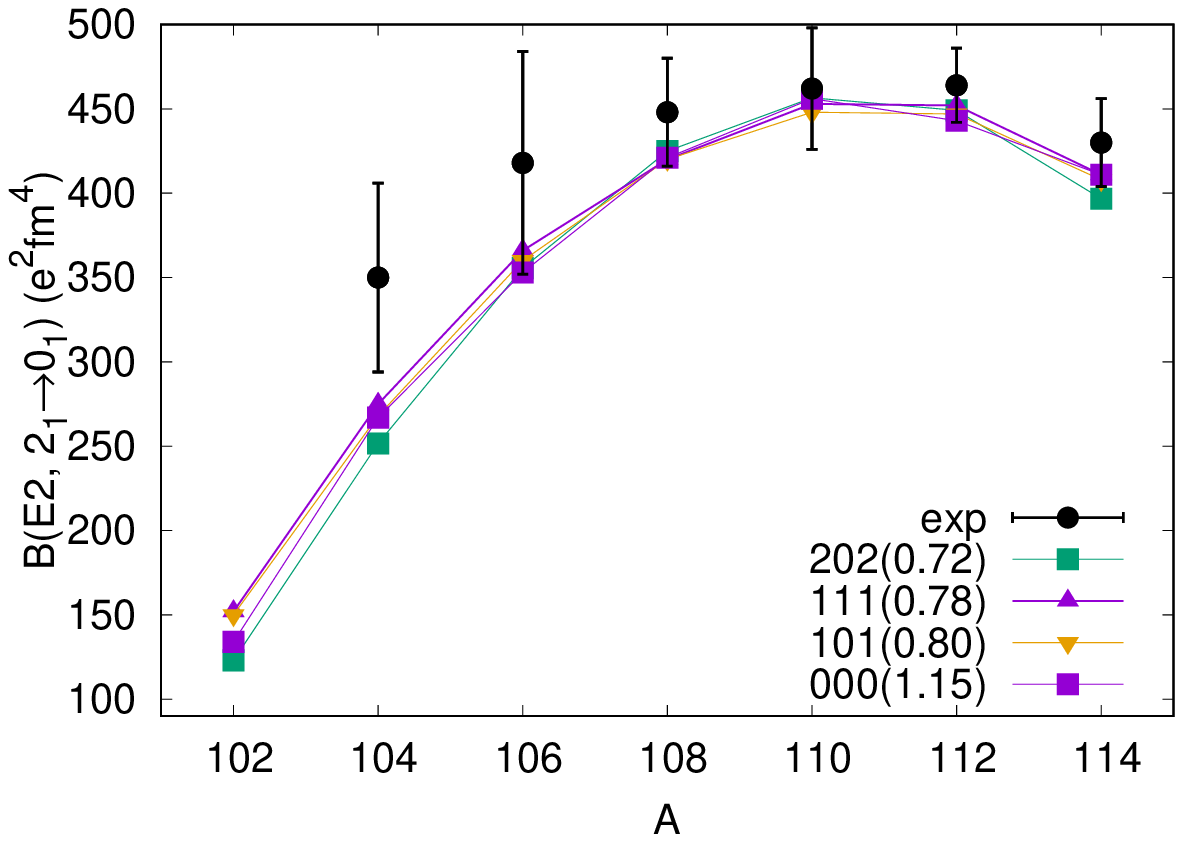}
\includegraphics[width=0.5\textwidth]{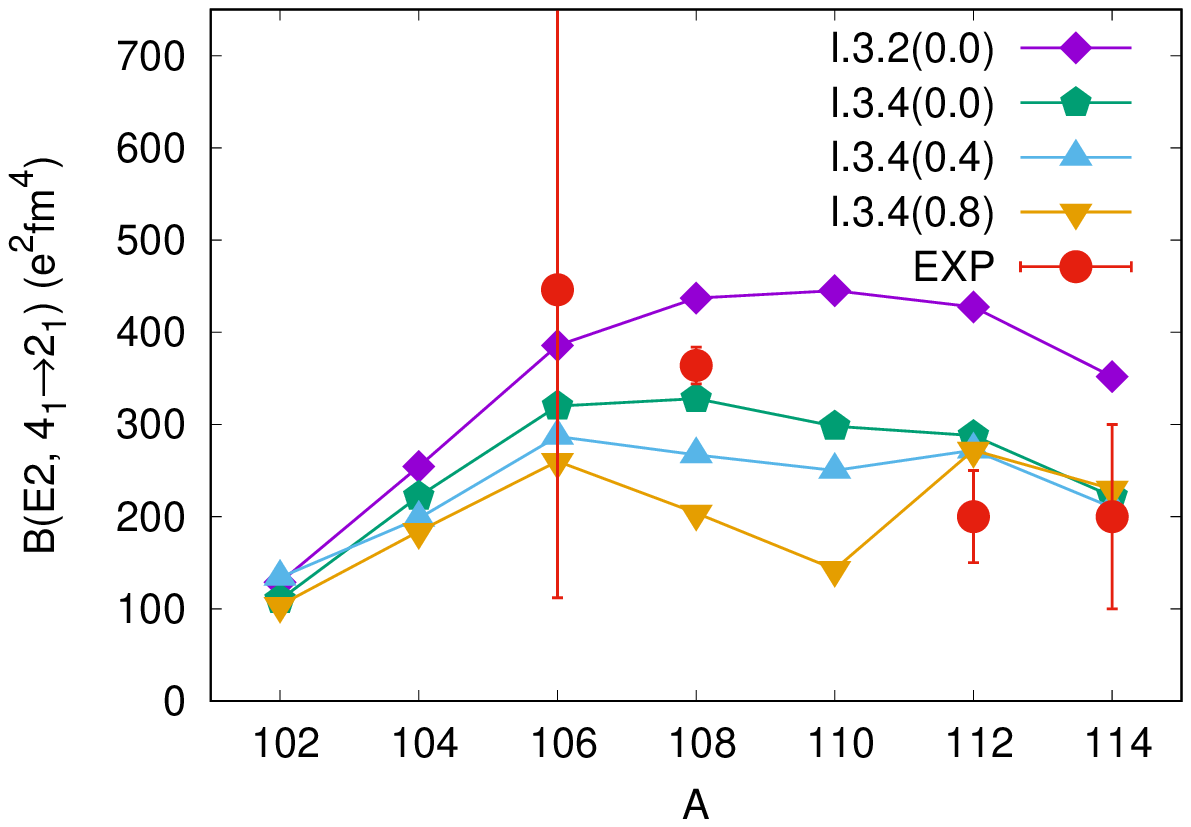}
\end{center}
\caption{\label{fig:be2042} Upper panel: Experimental~\cite{be22016}
  and calculated \bet values for Sn isotopes with the I.3.4
  interaction for different $utM$ values. In parenthesis $(e_{\nu})$,
  $e_{\pi}=1.40$ is fixed. Lower panel: \bef data from Jonsson
  \etal~\cite{jonsson1981} for \A{112-114}Sn, and from Siciliano
  \etal~\cite{siciliano2020} for \A{106-108}Sn. I.3.2($\delta$) and
  I.3.4($\delta$) are $utM=202$ calculations with the $s_{1/2}$
  single-particle energy displaced by $\delta=$0.0, 0.4 and 0.8 MeV
  with respect to the GEMO value of 0.8 MeV.}
\end{figure}

The basic tenet of this paper is that quadrupole dominance is
responsible for the \bet patterns in the light Cd and Sn
isotopes, which means that they should exhibit a pseudo-SU3
symmetry. Hence, an intrinsic state should exist,
implying the validity of the Alaga rule (Eq.~\eqref{bme2}). The
expectation is fulfilled in Cd (Fig.~\ref{fig:Cdbe2}) but it fails in
Sn, as seen in Fig.~\ref{fig:be2042}, where
the \bet rates are consistent with pseudo-SU3 validity, and are immune
to details, while the \bef rates are sensitive to the single-particle field
and to the pairing strength.
To test how the \bef rates are influenced by the single particle
field, the energy of the $s_{1/2}$ orbit in \A{101}Sn was displaced by
0.0, 0.4 and 0.8 MeV with respect to the present GEMO
choice~\cite{gemo}, called DZ (Duflo Zuker) in
Ref.~\cite[Fig. 3.2.1]{Sn100} where an extrapolated value (EX) is
given as reference. The position of the $s_{1/2}$ orbit for DZ and EX
differ by 800 keV. In the calculations reported in
Fig.~\ref{fig:be2042}, I.3.4(0,0) and I.3.4(0.8) correspond to DZ and
EX respectively.  The \bef differences are significant. Thanks to the
recent \A{108}Sn \bef measure of Siciliano
\etal~\cite[Fig. 4b]{siciliano2020}, the DZ choice is clearly favored.

\subsection{\label{sec:beqano}The \beq anomaly and the
  pairing-quadrupole interplay}
The \beq$<1$ anomaly had been detected in
\A{114}Xe~\cite{deangelis2002coherent}, in
\A{114}Te~\cite{moller2005e2} , and more recently in \A{172}Pt,
Ref.~\cite{cederwall2018}, where it is stressed that no theoretical
explanation is available.  Here, the sensitivity to the pairing
strength provides a clue in Fig.~\ref{fig:be2042}, where its decrease
in going from I.3.4 to I.3.2 produces a substantial increase of
\bef. The effect is seen most clearly in Ref.~\cite[Figs. 4a and
4b]{siciliano2020}, where \bet is seen to be totally immune to
pairing, while \bef is so sensitive that a sufficient decrease in
strength could bring \beq close to the Alaga rule. It appears that
pairing is eroding the deformed band. Only the lowest $J=0$ and 2 are
spared, giving way to a pairing-quadrupole interplay, that will
eventually end up in pairing dominance at $N\approx 70$. The
transition region will be studied in Sec.~\ref{sec:N>64}.

\section{The interaction and the model spaces}\label{sec:SM}
\begin{figure}[b]
  \begin{center}
 \includegraphics[width=0.5\textwidth]{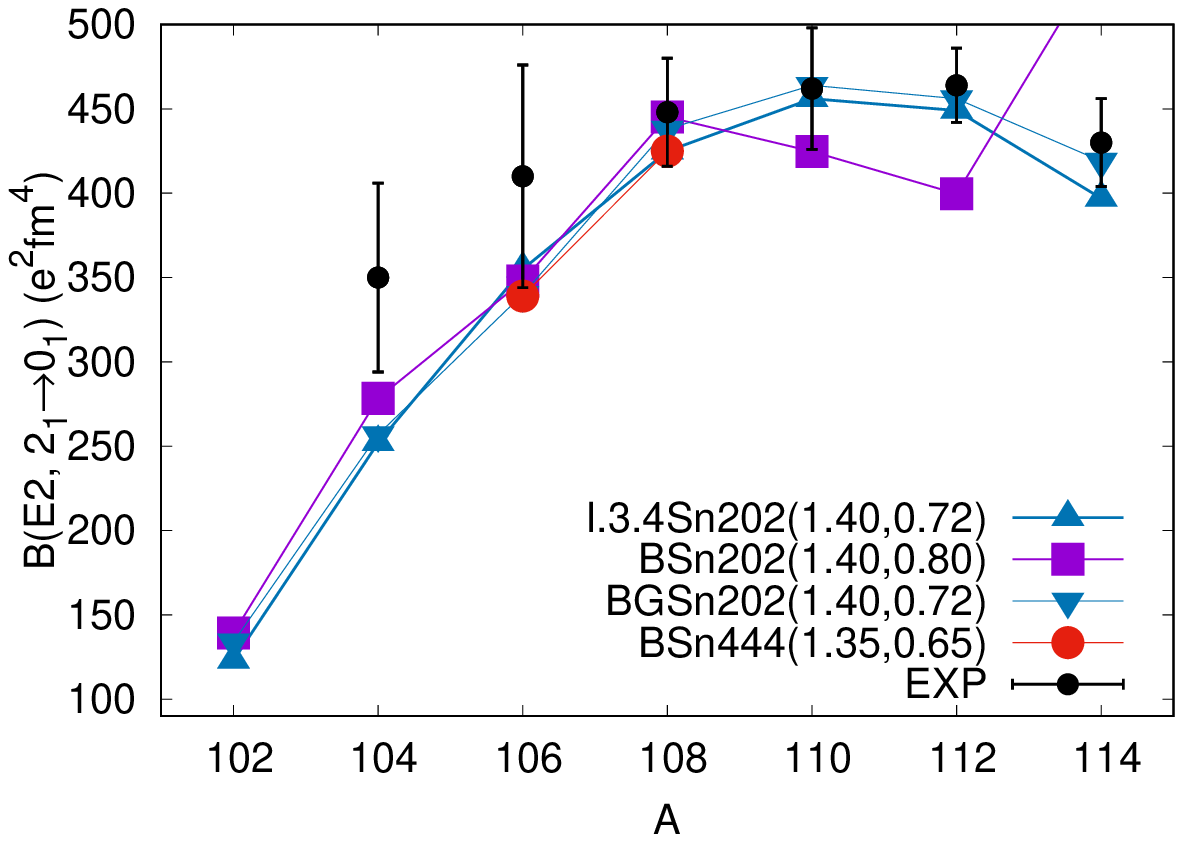}
\includegraphics[width=0.5\textwidth]{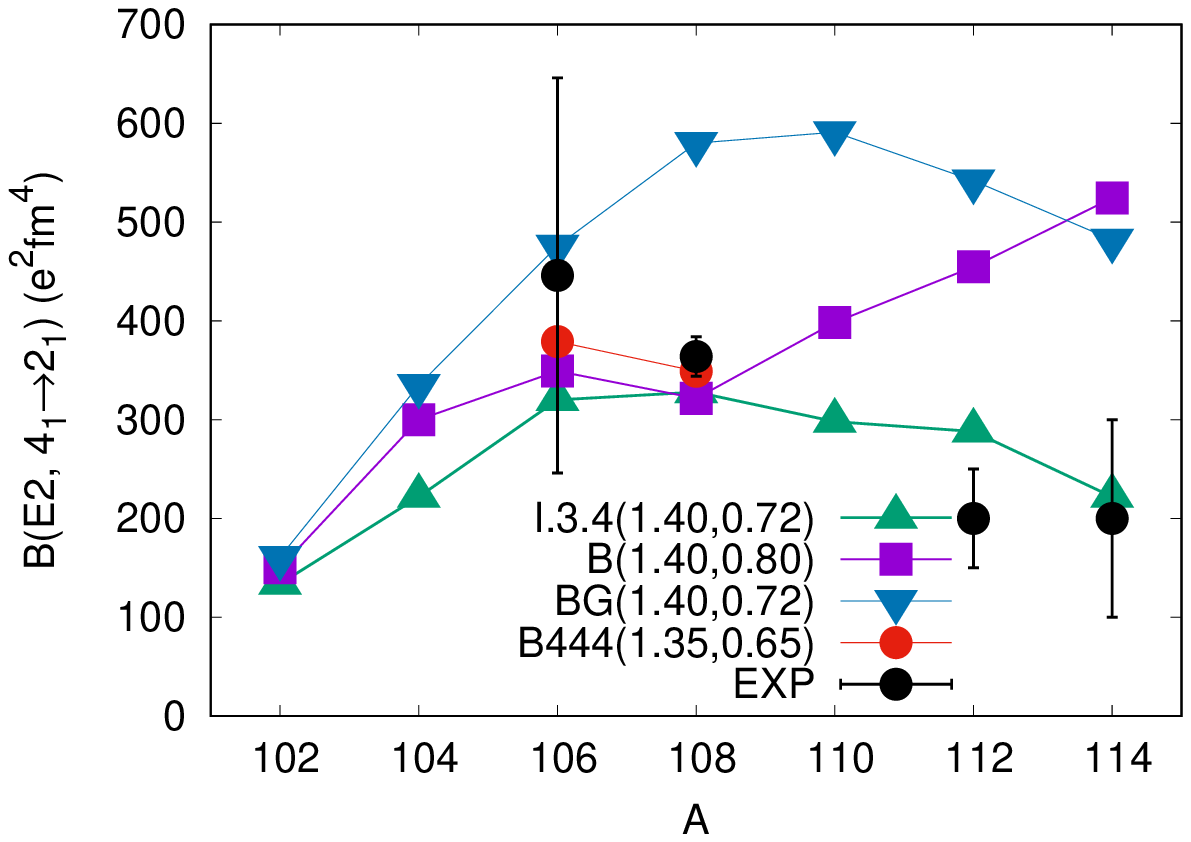}
\end{center}
\caption{\label{fig:comp} Comparing \bet for different $utM$ values
  (upper panel) and \bef (lower panel) calculations ($utM=202$) for B
  (the interaction used in Ref.~\cite{banu2005}, omitting the
  $h_{11/2}$ orbit) and I.3.4 interactions in the Sn isotopes.  BG is
  B made monopole free with the same GEMO single particle spectrum as
  I.3.4. In parenthesis \epinu. Experimental data as in
  Fig.~\ref{fig:be2042}.  See text. }
\end{figure}

Traditionally, the tin region was taken to be the prime example of
pairing dominance, and calculations were based on seniority
truncations which lead naturally to an overall
parabolic trend, as shown schematically in Fig.~\ref{fig:SnBE2}.

The \bet trend in the upper panel of Fig.~\ref{fig:be2042}, is already
well reproduced by the neutron-only ($utM=000$) case. It will be shown
that the monopole field is responsible for this welcome result. Though
B\"ack and coworkers~\cite[Fig. 3]{BE2Sn100} already suggested that in
such spaces the \bet trend could be altered by changes in the
single-particle behavior, their result was only indicative.  The more
complete calculations of Togashi
\etal~\cite[Fig. 2]{PhysRevLett.121.062501} did better quantitatively,
but $g$ proton excitations proved indispensable.  (No \bef estimates
were given in this reference).

The difficulty in reproducing the \hbox{\bet} enhancements for $A<114$
persists even when proton excitations are allowed in calculations
~\cite{banu2005,ekstrom2008sn,BE2Sn100} using the CDB (charge
dependent Bonn) potential~\cite{CDB}, renormalized following
Ref.~\cite{Hjorth-Jensen.Kuo.Osnes:1995} (B in what follows). This
raises two questions: why the I.3.4 interaction succeeds where others
fail?  and why the neutron-only description is not only viable, but
the correct basic model space? They can be answered simultaneously and
I start by explaining how severely truncated spaces may represent the
exact results, by comparing the largest calculation available with
smaller ones. In Ref.~\cite[Table I]{siciliano2020}, results are given
for \A{106-108}Sn in $utM=444$ ($m$-dimensions $10^{10}$) with the B
interaction used in Banu \etal~\cite{banu2005} (but omitting the
$h_{11/2}$ orbit).  In Fig.~\ref{fig:comp} they are shown as B444
(circles) and compared with B202 (squares, the same interaction in our
standard space).  The agreement is very good for the two points in
\bet and \bef. The result amounts to a splendid vindication of the
shell model viewed as the possibility to describe in a small space the
behavior of a large one. Although in general the reduction from large
to small spaces demands renormalization of the operators involved, for
our purpose, only the effective charges are affected, a non-trivial
fact that invites further study.

Let me return now to the crucial role of the monopole field.  In
Fig.~\ref{fig:comp}, the \bet discrepancies between I.3.4 and B in the
upper panel disappear when B is replaced by BG: the interaction made
monopole free and supplemented by the GEMO single-particle field used
in the I.q.p forces. For the \bef pattern in the lower panel, the
result is even more interesting: now BG is no longer close to I.3.4,
but to I.3.0, which is not shown, but can be guessed by extrapolation
in Fig.~\ref{fig:be2042} (from I.3.4 to I.3.2) and from the type of
multipole decomposition proposed in Ref.~\cite{mdz}, revealing the
same $q\cdot q$ content in I.3.4 and BG, and a much weaker pairing for
the latter; so weak in fact, that the BG results move closer to the
Alaga rule.
\begin{figure}[t]
  \begin{center}
\includegraphics[width=0.5\textwidth]{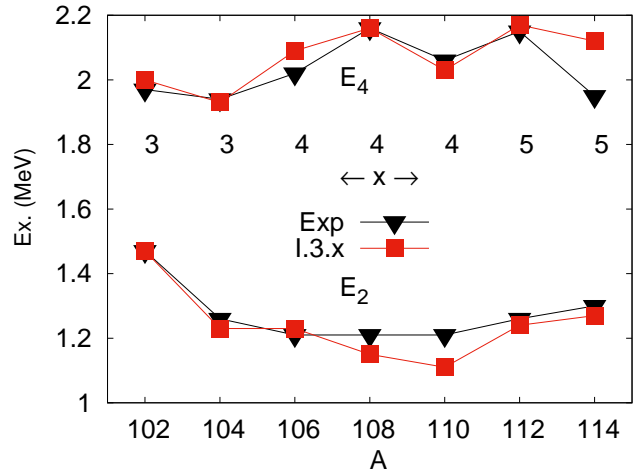}
\end{center}
\caption{\label{fig:EnSn} Lowest $E_2$ and $E_4$ energies, calculated
  with I.3.x interactions.}
\end{figure}

It follows that for I.3.0, say, the P$r_4$ symmetry will hold, at
least partially. As the pairing force is switched on, the $J=0_1,2_1$
states are not affected, while $J=4_1$ is. This points to an unusual
form of interplay between the two coupling schemes, pairing and
quadrupole, traditionally associated with collectivity. In single fluid
species, such as Sn, the seniority scheme can operate fully. It breaks
down in the presence of two kinds of particles, which turns out to be the
condition for quadrupole to operate successfully, as indicated by the
Cd isotopes. What is unusual is the presence of quadrupole coherence in
the light tins. It is shaky and challenged by pairing and it is known that
at about $A=120$ the seniority scheme will prevail.

For the transition nuclei \A{116-118}Sn, my original guess was that a
mixing of spherical and weakly oblate states would take place. I also
expected that neutron-only calculations, in which the monopole field
would play a crucial role, would be likely to shed light on these
matters. The guess was totally wrong, as explained in
Sec.~\ref{sec:prosn}. I mention this anecdote to stress that these two
nuclei were the hardest to understand in the region.

Throughout this study, energy spectra have been ignored in favor of
electromagnetic rates which are less sensitive to details, so I close
by showing spectra for the lowest $J=2$ and 4 levels. In
Fig.~\ref{fig:EnSn}, some pairing state dependence has been allowed,
but I.3.4 is seen to be the right overall choice.

\section{C\MakeLowercase{d} and S\MakeLowercase{n} at $\bf N \geq 64$}
\label{sec:N>64}

\begin{table}[t]
  \caption{\label{tab:Q2g}Intrinsic adimensional q0 for prolate (q0p),
and oblate (-q0o) states. Calculated spectroscopic quadrupole moments 
and g-factor\, Q2, Q4, g for I.3.4 \enupi=0.72, 1.40; $g_{s\nu}$=-2.869, 
$g_{l\nu}$=-0.070~\cite{neutron-gl}, $g_{s\pi}$=4.189, $g_{l\pi}$=1.100.           
Experimental Q2* and g* from Allmond \etal \cite{PhysRevC.92.041303}, 
$g_{s\nu\pi}$ quenched by 0.75 with respect to bare values~\cite[Fig.~28]{rmp}.}
\begin{ruledtabular}
\begin{tabular}{cccccccc}
N &  q0p&    -q0o&         Q2&  Q4&        Q2*&      g*   &    g  \\
52&   12&       6&        -18& -24&           &           & -0.157\\
54&   18&       12&       -21& -21&           &           &  0.012\\
56&   24&       18&       -16& -17&           &           &  0.103\\
58&   24&       24&        -5& -02&           &           &  0.142\\
60&   24&       24&         3&  10&           &           &  0.142\\
62&   24&       24&        14&  26&       4(9)&  0.150(43)&  0.135\\
64&   18&       24&        25&  43&       9(8)&  0.138(63)&  0.106\\
\end{tabular}
\end{ruledtabular}
\end{table}

In Table~\ref{tab:Q2g}, the naive P$r_4$ adimensional intrinsic
quadrupole moments for prolate (q0p) and oblate (q0o) are
compared. The former are the same as $q(n)_n$ in
Table.~\ref{tab:SP}. The latter are obtained by filling the platforms
in reverse order (from the top). Up to $N=56$ prolate dominates. From
$N=58$ to 62 there is oblate-prolate degeneracy. At $N=64$, oblate
dominates. In the absence of strong quadrupole dominance, these
intrinsic values only indicate a trend in sign, respected by the
calculated spectroscopic moments that opt for ``oblate'' shapes for
$A>108$. For \A{112-114}Sn, the shell-model results are close for the
quadrupole moments, or agree for the magnetic moments, with the
measured values. Note: The magnetic moments are very sensitive to the
anomalous $g_{l\nu}$.

By suggesting a very different behavior for the Sn and Cd and families
at N=64, that will be examined in what follows, Table~\ref{tab:Q2g}
illustrates the heuristic value of relying on pseudo-SU3, even when
the symmetry does not hold in the strict sense.

So far the $sdg$ space has proven sufficient, as the effects of the
$h_{11/2}$ orbit ($h$ for short) remain perturbative. For Sn, it is
known from classic $(p,d)$ work~\cite{CAVANAGH197097}, that the $h$
occupancy, very small up to \A{110}Sn, increases at \A{112-114}Sn,
as borne out by calculations that indicate the need of a boost of some
10\% in \bet~\cite{fred} with respect to Fig.~\ref{fig:be2042}. Beyond
$N=64$, the explicit inclusion of the $h$ orbit becomes imperative but
the situation is different for the two families. In Sn, it is known that 
the traditional $hr_4$ space will eventually prove sufficient when the
seniority scheme takes over at $N=70$. For \A{116-118}Sn, at this
stage, nothing can be said.  For Cd, the calculations give systematically 
prolate values in line with Stone's tables of quadrupole 
moments~\cite{STONE2016}, but in \A{112}Cd (excluded from both
Table~\ref{tab:SP} and Fig.~\ref{fig:Cdbe2}) they yield severe
underestimates whose correction necessitates the introduction of a
quasi-SU3 mechanism (referred generically as Q in what follows, and
Q$hfp$ for the case I introduced next).
It is illustrated in Fig.~\ref{fig:Cdx}, where it is seen that at
$N=64$, promoting an extra particle to the $q_0=-3$ platform reduces
prolate coherence in Cd, while filling the seven upper platforms makes
it possible for \A{114}Sn to stay oblate. To obtain realistic
estimates demands calculating the quadrupole moment in the presence of
a central field. An economic way of doing so is through Nilsson-SU3
self-consistency~\cite{nilssonSU3}, which is explained next.

\subsection{Nilsson-SU3 self-consistency in a SPQ context}

Let us start by remembering that $q_0=$\q, decompose it, together with
the corresponding operator $\hat q_0$, into the S, P and Q
contributions, and introduce the nornalized variant $\qn$.  Then
examining Ref.\cite[Eq.(19)]{nilssonSU3}
\begin{gather}
  \hat q_{0{\cal N}}=\frac{\hat q_{\pi S}}{{\cal N}_4} +\frac{\hat
    q_{\nu P}}{{\cal N}_4}+\frac{\hat q_{\nu Q}}{{\cal N}_5},
  q_{0{\cal N}}=\frac{ q_{\pi S}}{{\cal N}_4} +\frac{ q_{\nu
      P}}{{\cal N}_4}+\frac{q_{\nu Q}}{{\cal N}_5}\\
  H=H_{sp}-\frac{\hbar \omega \delta}{3}\hat{q_0}\equiv
  H_{sp}-\beta\hbar \omega \kappa
  \hat q_{0_{\cal N}}q_{0_{\cal N}}\label{hfin}\\
  {\cal N}^2=\sum (2q_{20rs})^2=\sum_{k=0}^p(k+1)(2p-3k)^2\label{N2}.
\end{gather}
($p$ is the principal quantum number).  Eq.~\eqref{hfin} compares the
classic Nilsson problem to the left and the self-consistent version to
the right, which demands the solution of a linearized
$\kappa \hbar \omega \hat{q_0} \cdot \hat{q_0}/{\cal N}^2$ problem, taken to
approximate Elliott's quadrupole force, in its correct realistic
normalized form, which involves the inclusion of the norm in
Eq.~\eqref{N2}, as demonstrated in Ref.~\cite{mdz}.  The coupling
constant, $\kappa=3$, is the same as in interactions I.3.x, while
$H_{sp}$ is taken from GEMO~\cite{gemo}. The quantity I am after,
$q_0$, is calculated while in the Nilsson case it is simply the
parameter $\delta$.

\begin{figure}[t]
  \begin{center}
\includegraphics[width=0.5\textwidth]{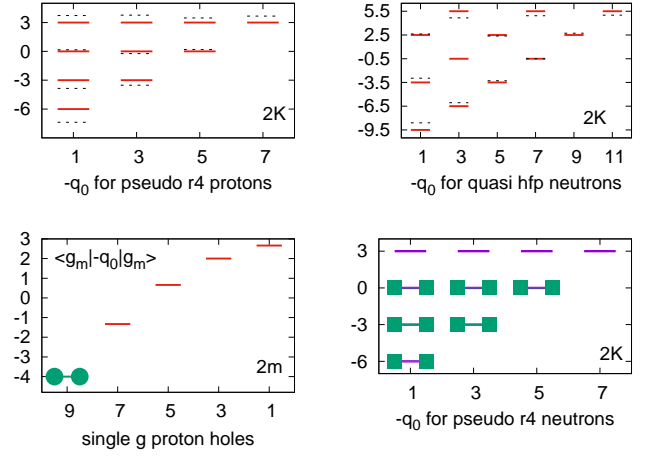}
\end{center}
\caption{\label{fig:Cdx} The \A{110}Cd intrinsic state in the SPQ 
space. The schematic pseudo-SU3 platforms P$r_{4s}$ are obtained by 
diagonalyzing the $\hat q_0$ operator in the $p=3$ space (full 
lines), while dashed lines correspond to the full $\hat q_0$ in the 
$r_4$ space (P$r_{4f}$). The quasi-SU3 (Q$hfp_s$) $q_0$ platforms
(full lines) are obtained by diagonalizing the quasi-quadrupole 
operator in the $hfp$ space \ie the degenerate $\Delta J=2$ sequence 
in $pfh$ shell: $h_{11/2}, \, f_{7/2}, \, p_{3/2}$. Dashed lines
(Q$hfp_f$) are for the full quadrupole operator. The quasi-quadruplole 
operator is obtained by using the $l\cdot s$ form of the $\hat q$ 
matrix elements and then replacing $l$ by 
$j=l+1/2$~\cite[Section IIIC]{nilssonSU3}.}
\end{figure}

The self-consistent solution of the problem is obtained by demanding
that that input and output $\qn$ coincide. Calculations are done for
each space separately. To ensure that the couplings involve the
full $q_0$, a parameter $\beta_X$ is introduced:

\begin{gather}
  \frac{\hat q_{\nu P}}{{\cal N}_4^2}( q_{\pi S}+ q_{\nu
    P})\longrightarrow \beta_P\frac{\hat q_{\nu P}}{{\cal
      N}_4^2}q_{\nu P}\label{betaP}\\
  \frac{\hat q_{\nu Q}}{{\cal N}^2_5}(q_{\nu Q}+(q_{\pi S}+q_{\nu
    P})\frac{{\cal N}_5}{{\cal N}_4})\longrightarrow \beta_Q
  \frac{\hat q_{\nu Q}}{{\cal N}^2_5}q_{\nu Q}\label{betaQ}
\end{gather}
At each iteration a full spectrum of Nilsson-like energies
$\varepsilon (2k,i)$ is generated, from which quadrupole contributions
$q_0(2k,i)$ are extracted by subtracting the $H_{sp}$ part. The full
$q_0$ is the sum of all such contributions for a given $A$. In the
case of \A{110}Cd in Fig.~\ref{fig:Cdx}, it involves the six filled
P-platforms. According to Eq.~\eqref{betaP}, this is the quantity to be
extracted self-consistently. However, it turns out that the results are
little changed if it is replaced by the single lowest contribution. In
other words, in \A{110}Cd $\beta_P$ may range from 1.3 for the full
$q_0=q_{\nu P}$ to 2.8 for $q_0(1,1)$, for nearly identical final
results $q_0=q_{\nu P}=$13.0(3), or 26.0(6) (for pair
occupancy) to be compared with 24 and 29.3, the values in Table I in
the absence of monopole field. Hence: the elementary SP arguments in
Table~\ref{tab:SP}, the diagonalizations in Fig.~\ref{fig:Cdbe2}, and
the present self-consistent results nearly coincide. This is a pleasant
result.

\begin{table}[t]
  \caption{\label{tab:Cdall} Quantities entering schematic and
    self-consistent calculations for \A{112-122}Cd ($N=64-74$). Single
    particle spectrum at \A{115}Sn from GEMO~\cite{gemo}.
    $\epsilon(h_{11/2})=0,\, \epsilon(f_{7/2})=3.6, \,
    \epsilon(p_{3/2})=5.4$ MeV. See text for detailed explanations.}
\begin{ruledtabular}  
  \begin{tabular}{ccccccc}
  \multicolumn{7}{c}{Qhfp$_f$ (Q$_f$)}\\
$2ki$&           11&       31&      51&      12&      71&      32 \\
$\phi^2_{h}$&0.21&    0.43&   0.69&   0.61&   0.89&    0.51\\
$q_{0}(2k,i)$&       8.55&     6.06&   3.28&   2.96&   0.50&    0.50\\
$N$ &           64&     66&    68&    70&   72&  74 \\
    $q_{\nu Q}(N)$ &8.55&    14.61&  17.89&  20.85&  21.35& 21.65\\
    $e_{\nu}(N)$    &0.75&     0.60&   0.55&  0.50&    0.50&  0.50\\
$B(E2,A)$& 1016&  1049&   1095&   1095&   1130&  1165\\
  \multicolumn{7}{c}{Self consistent (SC)}\\
$2ki$ &          11&        31&       51&      71&     91&     111\\
$\phi^2_{h}$&   0.97& 0.98&    0.99&   0.99&   1.00& 1.00 \\
$\varepsilon(2k,i)$&  -0.72& -0.59& -0.35& -0.01& 0.43& 0.94\\
$q_{0}(2k,i)$&  4.44& 3.62& 2.13& 0.13& -2.27& -5.00 \\
    $A$ &           112&     114&    116&    118&   120&  122 \\
    $q_{\nu Q}(A)$& 4.44& 8.06& 10.19& 10.32& 8.05& 3.05\\
$e_{\nu}(A)$    &0.95& 0.83& 0.78& 0.78& 0.75& 0.75\\
 $B(E2,A)$&984& 1078& 1148& 1169& 934& 655 \\
\end{tabular}
\end{ruledtabular}
\end{table}

At $N=64$ and beyond, Eq.~\eqref{betaQ} applies. Since
${\cal N}^2=210$ and 420 for $p=4$ and 5 respectively,
$(q_{\pi S}+q_{\nu P}){\cal N}_5/{\cal N}_4= (4+13)\sqrt 2\approx 24$.
The self-consistent calculations in Table~\ref{tab:Cdall} are done using
$q_0=q_0(1,1)+q_0(3,1)$ input-ouput values and $\beta_Q=4$, consistent
with $q_{\nu Q}=8.06$ in the table, as (24+8)/8=4.

In results so far, involving P and Q spaces (Refs.~\cite{Q} for
the rare earth, \cite{nilssonSU3} for $N=Z$ nuclei and for $N<64$ in
the present study) , the influence of $H_{sp}$ is relatively minor,
and \bet rates remain close to their theoretical maxima represented by
the $q_0$ diagrams. For the Q case, described in Table~\ref{tab:Cdall},
$H_{sp}$ plays a major role, and I have chosen to present together the
results for the strict Q$hfp_f$ and self-consistent cases (Q$_f$ and SC
in what follows).

Under $\phi^2_{h}$, I have listed the squared amplitude of the
$h=h_{11/2}$ components of the wave functions. They are on the average
of about 56\% for Q$_f$ and 98\% for SC, leading to different filling
patterns.

The filling patterns ($2ki$) as a function of $A$ are those of
Fig.~\ref{fig:Cdx} for Q$_f$, but are dictated by the energies
$\varepsilon(2k,i)$ in MeV for SC, in which case the $q_0(2k,i)$ values
are not necessarily the largest possible.

In particular, $q_0(1,2)\approx 5$ is the largest, but it has a huge
energy $\varepsilon(1,2)=2.82$ MeV, and $\phi^2_{h}\approx 0$. The
(12) orbit will play two roles in what follows: as a purveyor of
intruders and as signaling a transition between two deformed regimes.

\begin{figure}[t]
  \begin{center}
\includegraphics[width=0.5\textwidth]{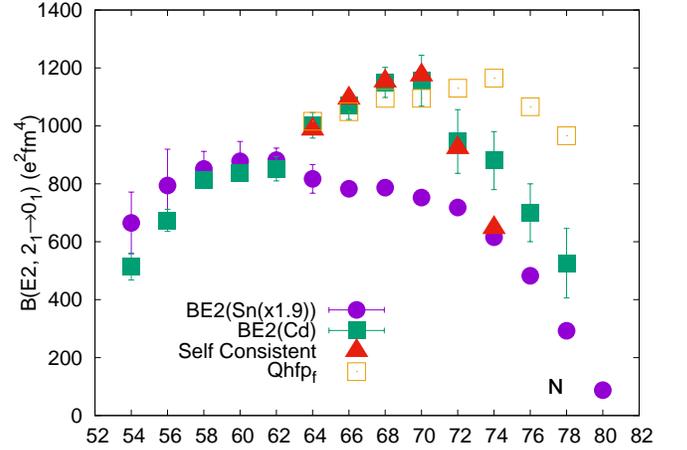}
\end{center}
\caption{\label{fig:CdSnall} The observed \bet values for Cd and Sn,
  and the schematic and self-consistent estimates for Cd at $N=64$ and
  beyond. Note that the Sn values are boosted by a 1.9 factor.}
\end{figure}

The $B(E2,A)$=\bet rates, with $e_{\nu}(A)$ from Table~\ref{tab:Cdall},
are calculated through
\begin{gather}\label{cd110be2}
 B(E2: 2_1\to 0_1) =[2(q_{\pi S}e_{\pi} +(q_{\nu P}+q_{\nu
   Q})e_{\nu})b^2]^2/50.3,
\end{gather}
where $q_{\pi S}=4,\, e_{\pi}=1.4$, and $q_{\nu P}=13$, as calculated
earlier. To account for rearrangements in the P space when Q pairs are
added $q_{\nu P}=14$ would also do, without affecting the satisfactory
agreement with data in Fig.~\ref{fig:CdSnall}.  To explain the
$e_{\nu}(A)$ choices , I start recalling that they are associated with
2\hw and 0\hw contributions, described in the paragraphs preceeding
Sec.~\ref{sec:cd}. Up to $N=64$, the 0\hw part is mediated by
quadrupole jumps coupling $r_4$ neutrons to proton particle-hole
excitations from the $g$ orbit. For $N\geq 64$ the couplings are
increasingly mediated by $hfp$, \ie $p=5$ neutrons, leading to a
suppression of quadrupole strength, due to the ${\cal N}_5/{\cal N}_4$
norm effect that has been encountered earlier. As a consequence
$e_{\nu}$ is expected to decrease gradually as the $r_4$ orbits fill
and reduce their contribution. The alternative is a constant,
plausible for $N>64$ but not at $N=64$, which definitely demands a
larger $e_{\nu}$. Once the gradual decrease is accepted, the choice in
Table~\ref{tab:Cdall} is quite constrained and the good agreement in
Fig.~\ref{fig:CdSnall} follows naturally. The table and
Eq.\eqref{cd110be2} contain all that is needed to explore
alternatives, but they will hardly change the quality of the
agreement.

Independently of details, there are two indications that the
calculations are on track: the drop at \A{120}Cd, $N=72$, and the
strong underestimate at \A{122}Cd which signals the transition to
oblate states, suggested by Stone's tables of quadrupole
moments~\cite{STONE2016}. The point is delicate because the signs
depend on the $K$ values (Eq.~\eqref{bmq}). The hint comes from the
$h_{11/2}$ values, systematically negative up to \A{119}Cd, and
positive beyond.

The two indications are probably correlated and invite further study.
 
\subsection{\label{sec:proscd}Cd prospects: vibration, intruders,
   coexistence}
 Arguably, the most striking feature of the self-consistent calculation
 is the overwhelming dominace of the $h=h_{11/2}$ orbit, which makes
 it impossible to speak of a quasi-SU3 symmetry, though one works in a
 Q space. The results of calculations in the full $p=5$ space, using
 $H_{sp}$ from GEMO~\cite{gemo}, are nearly identical to those in the
 Q space, which is vindicated as the correct choice. The transition
 between the Q$_f$ and SC regimes can be followed through variations of
 the $H_{sp}$ splittings. At
 $\epsilon(h_{11/2})=0,\, \epsilon(f_{7/2})=2.0, \,
 \epsilon(p_{3/2})=3.0$ MeV, $\phi^2_h=0.89$ and the SC filling
 pattern remains unchanged, but with a further reduction:
 $\epsilon(h_{11/2})=0,\, \epsilon(f_{7/2})=1.0, \,
 \epsilon(p_{3/2})=2.0$ MeV, $\phi^2_h=0.66$ the patterns change to
 Q$_f$, as the $2ki=12$ orbit fills at $N=72$: there is a change in
 regime from SC to Q$_f$. As to what is SC in Table~\ref{tab:Cdall}, an
 answer is suggested in the tables~\cite{nudat2}: the nuclei must be
 vibrational since \beq$\approx 2$ . The problem is that vibrational nuclei are
 not supposed to be deformed, as pointed out by Tamura and
 Udagawa~\cite{tamura1966} after the observation of a large static
 quadrupole moment in \A{114}Cd. After half a century, the question
 remains open~\cite{garrett2010,heyde2011}, though attempts have been
 made to modify the vibrational model so as to make it
 viable~\cite{leviatan2018}. As of now, I adopt a Gordian-knot solution
 and call such states q-vibrational. The precise definition will be
 given in Sec.~\ref{sec:qvib}.

 In introducing the SPQ spaces, I expected \A{112-116}Cd to behave as
 weakly deformed states in analogy to their lighter counterparts, but
 something different is happening, as made clear in the calculations.
 In \cite[Fig. 5]{garrett2010} the contrast is clear: \A{110}Cd
 follows the Alaga rule, while \beq= 2 for \A{112-116}Cd.  

 Recent beyond mean field (BMF) calculations in
 \A{110,112}Cd~\cite{garrett2019} are fairly successful at describing
 the numerous coexisting states. Referring to Fig.~\ref{fig:Cdx}, it is
 easy to visualize how such states could be produced by promoting
 $q_0=0$ P pairs to the Q$_f$ space. Thus, promoting the $2ki=12$ pair
 with $q_0(1,2)\approx 5$ on top of \A{112-114}Cd ground states leads
 to $q_{\nu Q}\approx 9$ and 13 respectively, using $e_{\nu}$ from
 Table~\ref{tab:Cdall}, and then \bett= 47 and 48 W.u., respectively,
 against the observed 51(12) and 65(9) W.u. The case of \A{110}Cd does
 not demand calculations but a check: its \bett=29(5) W.u. should be
 the same as \bet=30.3(2) in \A{112}Cd. They are. The experimental values
 are from~\cite{be22016,garrett2010,garrett2019}.

 The calculations and estimates so far, are useful in describing \bet
 trends but the determinant test rests with the \hbox{\beq} ratios.  In
 the case of the light Cd isotopes, it was met by shell-model
 calculations, but the irruption of the Q space puts them beyond reach
 for $N\ge 64 $. The BMF spectra are of little help, as they yield
 ratios too large compared to the Alaga rule in \A{110}Cd (1.66), and
 below the vibrational limit in \A{112}Cd (1.71)~\cite{garrett2019}.
 So, I propose to try something different.

 \subsubsection{Band Coupling}

 The idea is to prediagonalize the Hamiltonian in the SP and Q
 subspaces and couple the resulting bands to form a basis
$J_{SP}\bigotimes J_{Q}$.

In all probability, something similar has been proposed in the past,
but I know of no successful implementation, probably because of the
difficulty of defining the correct interaction to be used in the
individual spaces, as can be understood by concentrating on the
quadrupole force. The naive view is that, if $k\hat q\cdot \hat q$ is
used in the full space, then the same should be used in each of the
subspaces. What has been learned is that the correct choice is to
change $k\to \beta k$. Though no calculations in the coupled basis
will be attempted here, some runs were made in the Q or SQ spaces,
$hfp^{2-10}_{\nu}$ or $(h^{10}_{\pi})(hfp^{2-10}_{\nu})$ (replacing
$g^8_{\pi}$ by $h^{10}_{\pi}$, for simplicity) with a large quadrupole
force and with the realistic I interaction with the same quadrupole
strength. The results with both interactions very much coincide with
those of the SC calculations for \bet but nothing close enough to
\beq= 2 emerged. It is to be hoped that it will happen once the
coupling to the SP space is implemented. Still, the rudimentary
calculations confirmed that the quadrupole force determines the
coupling scheme, and that, even with very large $H_{sp}$ splittings,
and overwhelming $h_{11/2}$ dominance, prolate solutions prevailed,
even for two particles. I propose to call such states
$h^2$-prolate. They will prove important in what follows.

 \subsection{\label{sec:prosn} Sn prospects in A=116 and 118.}
 
Following Montaigne again, I start with a spoiler: \A{116-118}Sn are
 most probably q-vibrational nuclei. In spite of the spoiler, the story
 is of interest. At $N=64$, the interaction favors oblate in
 Table~\ref{tab:Q2g} (consistent with data in Allmond 
\etal~\cite{PhysRevC.92.041303}), and the calculations do well,
 (Fig.~\ref{fig:Cdbe2}). For Cd, there is a change in regime marked by
 a jump in Fig.~\ref{fig:CdSnall}. For Sn, there is a smooth inflection
 point, inviting the idea of a smooth transition. The natural
 assumption of an oblate \A{116}Sn, obtained by adding an $h^2_{11/2}$
 pair does not work for two reasons. The first is that the \bet of about
 700-800 \eff or 20-24 W.u. (adapting Eq.~\eqref{cd110be2},
 eliminating protons and replacing $q_{\nu Q}$ by $q_{\nu h}=-5$), is
 double the observed value. The second is that \A{116}Sn is
 prolate~\cite{STONE2016}. The way out is to use the prolate solution
 in Table~\ref{tab:Q2g}, \ie $q_{\nu P}$= 18 rather than -24, and
 $h^2$-prolate \ie $q_{\nu Q}=4.44$ from Table~\ref{tab:Cdall} in
 Eq.~\eqref{cd110be2} to obtain 12.1 W.u. for $e_{\nu}=1.05$, against
 the observed 12.4(4) W.u.

 All this may seem far fetched, but the corroborating evidence is strong:
as \bef= 38(24) W.u., the \beq quotient becomes
vibrational-compatible. For \A{118}Sn, the situation is similar.

\subsection{\label{sec:qvib}q-vibrations}

The total coincidence between Cd and Sn \bet patterns in
Fig.~\ref{fig:CdSnall} at $N<64$ was challenged by the \beq quotients,
which ironically establish their similarity at $N\ge 64$ where they
diverge abruptly. I have used the term q-vibrational for the latter
region. It applies to states that fulfill two conditions: a) \hbox{\beq}
and~$E_4/E_2\approx 2$, and b) sizable quadrupole moment of
non-rotational origin, as calculated in the previous
Secs.~\ref{sec:proscd} and ~\ref{sec:prosn}. Condition b ensures
a situation similar to that encountered at the beginning of this
study: the encouraging SP suggestions for \bet were validated by
further shell-model work, still missing here, but expected to work
equally well. 

It could be objected that postulating q-vibrations is a bold step
that relies too heavily on data. Certainly, but to explain why \beq is
smaller than one in \A{114}Sn and about two in \A{116}Sn one has to be
a bit bold. And remember that the speculations rest on credible \bet
estimates.

\section{Conclusions}
For a summary of what has been achieved in this paper, I refer to the
``Results'' item of the abstract, and concentrate here on the
basic ingredients at the origin of the good results: the interaction and
specifically, the monopole Hamltonian, and the SPQ interpretive
framework. 

\begin{description}
\item[The Interaction]The shell-model Hamiltonian can be separated
  into a monopole part, $H_m$, that contains all quadratic terms in
  number operators, and a multipole part, $H_M$, that contains all the
  rest, and is dominated by pairing and quadrupole components that
  demand a renormalization for use in restricted spaces. While $H_M$
  depends very little on the realistic potential chosen~\cite[II
  D]{rmp}, and does well spectroscopically, $H_m$, responsible for
  saturation properties and single-particle behavior, usually
  necessitates {\em ad hoc} fits, adapted to the problem at hand. In
  this work, I have relied on the general monopole Hamiltonian
  GEMO~\cite{gemo}, that reproduces, within a root mean square error
  of 200 keV, the particle and hole spectra on double-magic nuclei.
\item[The monopole interaction] A key step in the implementation is to
  eliminate the monopole part $H_m$ and use the GEMO
  single-particle field, eventually modulated by the total number of
  active particles; not necessary for $N<64$ but essential beyond,
  when the $h_{11/2}$ orbit comes in. The GEMO choice has been found
  to ensure correct \bet and \bef patterns in Figs. 3, 4 and 5, and has
  lead to a prediction for the position of the $s_{1/2}$ orbit in
  \A{101}Sn.

  Conceptually, GEMO has made it possible to establish the
  neutron-only description in the tin isotopes as the natural starting
  model space.
\item[The SPQ interpretive framework and self consistency] In doing
  calculations, the aim is not only to reproduce data, but also to
  understand underlying mechanisms. The SPQ representations, in
  addition to their predictive power, have served this purpose
  well. They are also the backbone of Nilsson-SU3 selfconsistent
  estimates that supplement existing calculations and open the way to
  future ones, as in Fig.~\ref{fig:CdSnall}.
\end{description}  
In this paper, calculations have alternated between full rigor,
heuristics and semi-quantitative estimates, regions have moved from
well-developed deformed, to pairing-quadrupole coexistence, to the
newly postulated q-vibrational states. Cadmium and tin come and go.  ``The
world is but a perennial swing'' (Le monde n'est qu'une branloire
perenne. Essais III 2~\cite[p. 582]{montaigne}).

\begin{acknowledgments}
  Alfredo Poves and Fr\'ed\'eric Nowacki took an active interest in the
paper and made important suggestions.The collaboration with Marco
Siciliano, Alain Goasduff and Jos\'e Javier Valiente Dob\'on is
gratefully acknowledged.
\end{acknowledgments}

\bibliography{LE17749C}

\begin{thebibliography}{49}%
\makeatletter
\providecommand \@ifxundefined [1]{%
 \@ifx{#1\undefined}
}%
\providecommand \@ifnum [1]{%
 \ifnum #1\expandafter \@firstoftwo
 \else \expandafter \@secondoftwo
 \fi
}%
\providecommand \@ifx [1]{%
 \ifx #1\expandafter \@firstoftwo
 \else \expandafter \@secondoftwo
 \fi
}%
\providecommand \natexlab [1]{#1}%
\providecommand \enquote  [1]{``#1''}%
\providecommand \bibnamefont  [1]{#1}%
\providecommand \bibfnamefont [1]{#1}%
\providecommand \citenamefont [1]{#1}%
\providecommand \href@noop [0]{\@secondoftwo}%
\providecommand \href [0]{\begingroup \@sanitize@url \@href}%
\providecommand \@href[1]{\@@startlink{#1}\@@href}%
\providecommand \@@href[1]{\endgroup#1\@@endlink}%
\providecommand \@sanitize@url [0]{\catcode `\\12\catcode `\$12\catcode
  `\&12\catcode `\#12\catcode `\^12\catcode `\_12\catcode `\%12\relax}%
\providecommand \@@startlink[1]{}%
\providecommand \@@endlink[0]{}%
\providecommand \url  [0]{\begingroup\@sanitize@url \@url }%
\providecommand \@url [1]{\endgroup\@href {#1}{\urlprefix }}%
\providecommand \urlprefix  [0]{URL }%
\providecommand \Eprint [0]{\href }%
\providecommand \doibase [0]{https://doi.org/}%
\providecommand \selectlanguage [0]{\@gobble}%
\providecommand \bibinfo  [0]{\@secondoftwo}%
\providecommand \bibfield  [0]{\@secondoftwo}%
\providecommand \translation [1]{[#1]}%
\providecommand \BibitemOpen [0]{}%
\providecommand \bibitemStop [0]{}%
\providecommand \bibitemNoStop [0]{.\EOS\space}%
\providecommand \EOS [0]{\spacefactor3000\relax}%
\providecommand \BibitemShut  [1]{\csname bibitem#1\endcsname}%
\let\auto@bib@innerbib\@empty
\bibitem [{\citenamefont {Pritychenko}\ \emph {et~al.}(2016)\citenamefont
  {Pritychenko}, \citenamefont {Birch}, \citenamefont {Singh},\ and\
  \citenamefont {Horoi}}]{be22016}%
  \BibitemOpen
  \bibfield  {author} {\bibinfo {author} {\bibfnamefont {B.}~\bibnamefont
  {Pritychenko}}, \bibinfo {author} {\bibfnamefont {M.}~\bibnamefont {Birch}},
  \bibinfo {author} {\bibfnamefont {B.}~\bibnamefont {Singh}},\ and\ \bibinfo
  {author} {\bibfnamefont {M.}~\bibnamefont {Horoi}},\ }\bibfield  {title}
  {\bibinfo {title} {Tables of e2 transition probabilities from the first $2^+$
  states in even-even nuclei},\ }\href@noop {} {\bibfield  {journal} {\bibinfo
  {journal} {At. Data Nucl. Data Tables}\ }\textbf {\bibinfo {volume} {107}},\
  \bibinfo {pages} {1} (\bibinfo {year} {2016})},\ \bibinfo {note} {~erratum
  \textit{ibid.} {\bf 114}, 371 (2017)}\BibitemShut {NoStop}%
\bibitem [{\citenamefont {Jonsson}\ \emph {et~al.}(1981)\citenamefont
  {Jonsson}, \citenamefont {B{\"{a}}cklin}, \citenamefont {Kantele},
  \citenamefont {Julin}, \citenamefont {Luontama},\ and\ \citenamefont
  {Passoja}}]{jonsson1981}%
  \BibitemOpen
  \bibfield  {author} {\bibinfo {author} {\bibfnamefont {N.-G.}\ \bibnamefont
  {Jonsson}}, \bibinfo {author} {\bibfnamefont {A.}~\bibnamefont
  {B{\"{a}}cklin}}, \bibinfo {author} {\bibfnamefont {J.}~\bibnamefont
  {Kantele}}, \bibinfo {author} {\bibfnamefont {R.}~\bibnamefont {Julin}},
  \bibinfo {author} {\bibfnamefont {M.}~\bibnamefont {Luontama}},\ and\
  \bibinfo {author} {\bibfnamefont {A.}~\bibnamefont {Passoja}},\ }\bibfield
  {title} {\bibinfo {title} {Collective states in even \textrm{Sn} nuclei},\
  }\href@noop {} {\bibfield  {journal} {\bibinfo  {journal} {Nucl. Phys. A}\
  }\textbf {\bibinfo {volume} {371}},\ \bibinfo {pages} {333} (\bibinfo {year}
  {1981})}\BibitemShut {NoStop}%
\bibitem [{\citenamefont {Banu}\ \emph {et~al.}(2005)\citenamefont {Banu},
  \citenamefont {Gerl}, \citenamefont {Fahlander}, \citenamefont {G\'orska},
  \citenamefont {Grawe}, \citenamefont {Saito}, \citenamefont {Wollersheim},
  \citenamefont {Caurier}, \citenamefont {Engeland}, \citenamefont {Gniady},
  \citenamefont {Hjorth-Jensen},\ and\ \citenamefont {Nowacki}}]{banu2005}%
  \BibitemOpen
  \bibfield  {author} {\bibinfo {author} {\bibfnamefont {A.}~\bibnamefont
  {Banu}}, \bibinfo {author} {\bibfnamefont {J.}~\bibnamefont {Gerl}}, \bibinfo
  {author} {\bibfnamefont {C.}~\bibnamefont {Fahlander}}, \bibinfo {author}
  {\bibfnamefont {M.}~\bibnamefont {G\'orska}}, \bibinfo {author}
  {\bibfnamefont {H.}~\bibnamefont {Grawe}}, \bibinfo {author} {\bibfnamefont
  {T.~R.}\ \bibnamefont {Saito}}, \bibinfo {author} {\bibfnamefont {H.-J.}\
  \bibnamefont {Wollersheim}}, \bibinfo {author} {\bibfnamefont
  {E.}~\bibnamefont {Caurier}}, \bibinfo {author} {\bibfnamefont
  {T.}~\bibnamefont {Engeland}}, \bibinfo {author} {\bibfnamefont
  {A.}~\bibnamefont {Gniady}}, \bibinfo {author} {\bibfnamefont
  {M.}~\bibnamefont {Hjorth-Jensen}},\ and\ \bibinfo {author} {\bibfnamefont
  {F.}~\bibnamefont {Nowacki}},\ }\bibfield  {title} {\bibinfo {title}
  {\textrm{$^{108}$Sn studied with intermediate-energy Coulomb excitation}},\
  }\href@noop {} {\bibfield  {journal} {\bibinfo  {journal} {Phys. Rev. C}\
  }\textbf {\bibinfo {volume} {72}},\ \bibinfo {pages} {061305(R)} (\bibinfo
  {year} {2005})}\BibitemShut {NoStop}%
\bibitem [{\citenamefont {Vaman}\ \emph {et~al.}(2007)\citenamefont {Vaman}
  \emph {et~al.}}]{vaman2007sn}%
  \BibitemOpen
  \bibfield  {author} {\bibinfo {author} {\bibfnamefont {C.}~\bibnamefont
  {Vaman}} \emph {et~al.},\ }\bibfield  {title} {\bibinfo {title}
  {\textrm{$Z=50$ shell gap near $^{100}$Sn from intermediate-energy Coulomb
  excitations in even-mass $^{106−112}$Sn isotopes}},\ }\href@noop {}
  {\bibfield  {journal} {\bibinfo  {journal} {Phys. Rev. Lett.}\ }\textbf
  {\bibinfo {volume} {99}},\ \bibinfo {pages} {162501} (\bibinfo {year}
  {2007})}\BibitemShut {NoStop}%
\bibitem [{\citenamefont {Ekstr{\"o}m}\ \emph {et~al.}(2007)\citenamefont
  {Ekstr{\"o}m}, \citenamefont {Cederk{\"a}ll}, \citenamefont {Fahlander},
  \citenamefont {Hurst}, \citenamefont {Hjorth-Jensen} \emph
  {et~al.}}]{cederkall2007sub}%
  \BibitemOpen
  \bibfield  {author} {\bibinfo {author} {\bibfnamefont {A.}~\bibnamefont
  {Ekstr{\"o}m}}, \bibinfo {author} {\bibfnamefont {J.}~\bibnamefont
  {Cederk{\"a}ll}}, \bibinfo {author} {\bibfnamefont {C.}~\bibnamefont
  {Fahlander}}, \bibinfo {author} {\bibfnamefont {A.}~\bibnamefont {Hurst}},
  \bibinfo {author} {\bibfnamefont {M.}~\bibnamefont {Hjorth-Jensen}}, \emph
  {et~al.},\ }\bibfield  {title} {\bibinfo {title} {\textrm{Sub-barrier Coulomb
  excitation of $^{110}$Sn and its implications for the $^{100}$Sn shell
  closure}},\ }\href@noop {} {\bibfield  {journal} {\bibinfo  {journal} {Phys.
  Rev. Lett.}\ }\textbf {\bibinfo {volume} {98}},\ \bibinfo {pages} {172501}
  (\bibinfo {year} {2007})}\BibitemShut {NoStop}%
\bibitem [{\citenamefont {Ekstr{\"o}m}\ \emph {et~al.}(2008)\citenamefont
  {Ekstr{\"o}m}, \citenamefont {Cederk{\"a}ll}, \citenamefont {Fahlander},
  \citenamefont {Hjorth-Jensen} \emph {et~al.}}]{ekstrom2008sn}%
  \BibitemOpen
  \bibfield  {author} {\bibinfo {author} {\bibfnamefont {A.}~\bibnamefont
  {Ekstr{\"o}m}}, \bibinfo {author} {\bibfnamefont {J.}~\bibnamefont
  {Cederk{\"a}ll}}, \bibinfo {author} {\bibfnamefont {C.}~\bibnamefont
  {Fahlander}}, \bibinfo {author} {\bibfnamefont {M.}~\bibnamefont
  {Hjorth-Jensen}}, \emph {et~al.},\ }\bibfield  {title} {\bibinfo {title}
  {\textrm{$0^+_{gs}\to 2^+_{1}$ transition strengths in $^{106}$Sn and
  $^{108}$Sn}},\ }\href@noop {} {\bibfield  {journal} {\bibinfo  {journal}
  {Phys. Rev. Lett.}\ }\textbf {\bibinfo {volume} {101}},\ \bibinfo {pages}
  {012502} (\bibinfo {year} {2008})}\BibitemShut {NoStop}%
\bibitem [{\citenamefont {Kumar}\ \emph {et~al.}(2010)\citenamefont {Kumar},
  \citenamefont {Doornenbal}, \citenamefont {Jhingan}, \citenamefont {Bhowmik},
  \citenamefont {Muralithar} \emph {et~al.}}]{kumar2010enhanced}%
  \BibitemOpen
  \bibfield  {author} {\bibinfo {author} {\bibfnamefont {R.}~\bibnamefont
  {Kumar}}, \bibinfo {author} {\bibfnamefont {P.}~\bibnamefont {Doornenbal}},
  \bibinfo {author} {\bibfnamefont {A.}~\bibnamefont {Jhingan}}, \bibinfo
  {author} {\bibfnamefont {R.~K.}\ \bibnamefont {Bhowmik}}, \bibinfo {author}
  {\bibfnamefont {S.}~\bibnamefont {Muralithar}}, \emph {et~al.},\ }\bibfield
  {title} {\bibinfo {title} {\textrm{Enhanced $0_{g.s.}^+ \rightarrow 2_1^+$
  $E2$ transition strength in $^{112}$Sn}},\ }\href@noop {} {\bibfield
  {journal} {\bibinfo  {journal} {Physical Review C}\ }\textbf {\bibinfo
  {volume} {81}},\ \bibinfo {pages} {024306} (\bibinfo {year}
  {2010})}\BibitemShut {NoStop}%
\bibitem [{\citenamefont {Bader}\ \emph {et~al.}(2013)\citenamefont {Bader},
  \citenamefont {Gade}, \citenamefont {Weisshaar} \emph
  {et~al.}}]{bader2013quadrupole}%
  \BibitemOpen
  \bibfield  {author} {\bibinfo {author} {\bibfnamefont {V.~M.}\ \bibnamefont
  {Bader}}, \bibinfo {author} {\bibfnamefont {A.}~\bibnamefont {Gade}},
  \bibinfo {author} {\bibfnamefont {D.}~\bibnamefont {Weisshaar}}, \emph
  {et~al.},\ }\bibfield  {title} {\bibinfo {title} {\textrm{Quadrupole
  collectivity in neutron-deficient Sn nuclei: $^{104}$Sn and the role of
  proton excitations}},\ }\href@noop {} {\bibfield  {journal} {\bibinfo
  {journal} {Phys. Rev. C}\ }\textbf {\bibinfo {volume} {88}},\ \bibinfo
  {pages} {051301(R)} (\bibinfo {year} {2013})}\BibitemShut {NoStop}%
\bibitem [{\citenamefont {Doornenbal}\ \emph {et~al.}(2014)\citenamefont
  {Doornenbal}, \citenamefont {Takeuchi}, \citenamefont {Aoi}, \citenamefont
  {Matsushita}, \citenamefont {Obertelli}, \citenamefont {Steppenbeck},
  \citenamefont {Wang} \emph {et~al.}}]{doornenbal2014intermediate}%
  \BibitemOpen
  \bibfield  {author} {\bibinfo {author} {\bibfnamefont {P.}~\bibnamefont
  {Doornenbal}}, \bibinfo {author} {\bibfnamefont {S.}~\bibnamefont
  {Takeuchi}}, \bibinfo {author} {\bibfnamefont {N.}~\bibnamefont {Aoi}},
  \bibinfo {author} {\bibfnamefont {M.}~\bibnamefont {Matsushita}}, \bibinfo
  {author} {\bibfnamefont {A.}~\bibnamefont {Obertelli}}, \bibinfo {author}
  {\bibfnamefont {D.}~\bibnamefont {Steppenbeck}}, \bibinfo {author}
  {\bibfnamefont {H.}~\bibnamefont {Wang}}, \emph {et~al.},\ }\bibfield
  {title} {\bibinfo {title} {\textrm{Intermediate-energy Coulomb excitation of
  $^{104}$Sn: Moderate E2 strength decrease approaching $^{100}$Sn}},\
  }\href@noop {} {\bibfield  {journal} {\bibinfo  {journal} {Phys. Rev. C}\
  }\textbf {\bibinfo {volume} {90}},\ \bibinfo {pages} {061302(R)} (\bibinfo
  {year} {2014})}\BibitemShut {NoStop}%
\bibitem [{\citenamefont {Kumar}\ \emph {et~al.}(2017)\citenamefont {Kumar},
  \citenamefont {Saxena}, \citenamefont {Doornenbal}, \citenamefont {Jhingan}
  \emph {et~al.}}]{kumar2017noevidence}%
  \BibitemOpen
  \bibfield  {author} {\bibinfo {author} {\bibfnamefont {R.}~\bibnamefont
  {Kumar}}, \bibinfo {author} {\bibfnamefont {M.}~\bibnamefont {Saxena}},
  \bibinfo {author} {\bibfnamefont {P.}~\bibnamefont {Doornenbal}}, \bibinfo
  {author} {\bibfnamefont {A.}~\bibnamefont {Jhingan}}, \emph {et~al.},\
  }\bibfield  {title} {\bibinfo {title} {\textrm{No evidence of reduced
  collectivity in Coulomb-excited Sn isotopes}},\ }\href@noop {} {\bibfield
  {journal} {\bibinfo  {journal} {Phys. Rev. C}\ }\textbf {\bibinfo {volume}
  {96}},\ \bibinfo {pages} {054318} (\bibinfo {year} {2017})}\BibitemShut
  {NoStop}%
\bibitem [{\citenamefont {Demonet}\ \emph {et~al.}(2016)\citenamefont
  {Demonet}, \citenamefont {Legros}, \citenamefont {Duboc}, \citenamefont
  {Bertrand},\ and\ \citenamefont {Lavrentiev}}]{montaigne}%
  \BibitemOpen
  \bibfield  {author} {\bibinfo {author} {\bibfnamefont {M.-L.}\ \bibnamefont
  {Demonet}}, \bibinfo {author} {\bibfnamefont {A.}~\bibnamefont {Legros}},
  \bibinfo {author} {\bibfnamefont {M.}~\bibnamefont {Duboc}}, \bibinfo
  {author} {\bibfnamefont {L.}~\bibnamefont {Bertrand}},\ and\ \bibinfo
  {author} {\bibfnamefont {A.}~\bibnamefont {Lavrentiev}},\ }\href@noop {}
  {\emph {\bibinfo {title} {{Michel de Montaigne, Essais, 1588 (Exemplaire de
  Bordeaux), {\'e}dition num{\'e}rique g{\'e}n{\'e}tique (XML-TEI/ PDF)}}}},\
  edited by\ \bibinfo {editor} {\bibfnamefont {M.-L.}\ \bibnamefont {Demonet}}\
  (\bibinfo {year} {2016})\BibitemShut {NoStop}%
\bibitem [{\citenamefont {Elliott}(1958{\natexlab{a}})}]{elliotta}%
  \BibitemOpen
  \bibfield  {author} {\bibinfo {author} {\bibfnamefont {J.~P.}\ \bibnamefont
  {Elliott}},\ }\bibfield  {title} {\bibinfo {title} {\textrm{Collective motion
  in the nuclear shell model. I. Classification schemes for states of mixed
  configurations}},\ }\href@noop {} {\bibfield  {journal} {\bibinfo  {journal}
  {Proc. R. Soc. London}\ }\textbf {\bibinfo {volume} {245}},\ \bibinfo {pages}
  {128} (\bibinfo {year} {1958}{\natexlab{a}})}\BibitemShut {NoStop}%
\bibitem [{\citenamefont {Elliott}(1958{\natexlab{b}})}]{elliottb}%
  \BibitemOpen
  \bibfield  {author} {\bibinfo {author} {\bibfnamefont {J.~P.}\ \bibnamefont
  {Elliott}},\ }\bibfield  {title} {\bibinfo {title} {\textrm{Collective motion
  in the nuclear shell model II. The introduction of intrinsic
  wave-functions}},\ }\href@noop {} {\bibfield  {journal} {\bibinfo  {journal}
  {Proc. R. Soc. London}\ }\textbf {\bibinfo {volume} {245}},\ \bibinfo {pages}
  {562} (\bibinfo {year} {1958}{\natexlab{b}})}\BibitemShut {NoStop}%
\bibitem [{\citenamefont {Zuker}\ \emph {et~al.}(1995)\citenamefont {Zuker},
  \citenamefont {Retamosa}, \citenamefont {Poves},\ and\ \citenamefont
  {Caurier}}]{Q}%
  \BibitemOpen
  \bibfield  {author} {\bibinfo {author} {\bibfnamefont {A.~P.}\ \bibnamefont
  {Zuker}}, \bibinfo {author} {\bibfnamefont {J.}~\bibnamefont {Retamosa}},
  \bibinfo {author} {\bibfnamefont {A.}~\bibnamefont {Poves}},\ and\ \bibinfo
  {author} {\bibfnamefont {E.}~\bibnamefont {Caurier}},\ }\bibfield  {title}
  {\bibinfo {title} {Spherical shell model description of rotational motion},\
  }\href@noop {} {\bibfield  {journal} {\bibinfo  {journal} {Phys. Rev. C}\
  }\textbf {\bibinfo {volume} {52}},\ \bibinfo {pages} {R1741} (\bibinfo {year}
  {1995})}\BibitemShut {NoStop}%
\bibitem [{\citenamefont {Caurier}\ \emph {et~al.}(2005)\citenamefont
  {Caurier}, \citenamefont {Mart\'{\i}nez-Pinedo}, \citenamefont {Nowacki},
  \citenamefont {Poves},\ and\ \citenamefont {Zuker}}]{rmp}%
  \BibitemOpen
  \bibfield  {author} {\bibinfo {author} {\bibfnamefont {E.}~\bibnamefont
  {Caurier}}, \bibinfo {author} {\bibfnamefont {G.}~\bibnamefont
  {Mart\'{\i}nez-Pinedo}}, \bibinfo {author} {\bibfnamefont {F.}~\bibnamefont
  {Nowacki}}, \bibinfo {author} {\bibfnamefont {A.}~\bibnamefont {Poves}},\
  and\ \bibinfo {author} {\bibfnamefont {A.~P.}\ \bibnamefont {Zuker}},\
  }\bibfield  {title} {\bibinfo {title} {The shell model as a unified view of
  nuclear structure},\ }\href@noop {} {\bibfield  {journal} {\bibinfo
  {journal} {Rev. Mod. Phys.}\ }\textbf {\bibinfo {volume} {77}},\ \bibinfo
  {pages} {427} (\bibinfo {year} {2005})}\BibitemShut {NoStop}%
\bibitem [{\citenamefont {Zuker}\ \emph {et~al.}(2015)\citenamefont {Zuker},
  \citenamefont {Poves}, \citenamefont {Nowacki},\ and\ \citenamefont
  {Lenzi}}]{nilssonSU3}%
  \BibitemOpen
  \bibfield  {author} {\bibinfo {author} {\bibfnamefont {A.~P.}\ \bibnamefont
  {Zuker}}, \bibinfo {author} {\bibfnamefont {A.}~\bibnamefont {Poves}},
  \bibinfo {author} {\bibfnamefont {F.}~\bibnamefont {Nowacki}},\ and\ \bibinfo
  {author} {\bibfnamefont {S.~M.}\ \bibnamefont {Lenzi}},\ }\bibfield  {title}
  {\bibinfo {title} {\textrm{Nilsson-SU3 self-consistency in heavy $N=Z$
  nuclei}},\ }\href@noop {} {\bibfield  {journal} {\bibinfo  {journal} {Phys.
  Rev. C}\ }\textbf {\bibinfo {volume} {92}},\ \bibinfo {pages} {024320}
  (\bibinfo {year} {2015})}\BibitemShut {NoStop}%
\bibitem [{\citenamefont {Arima}\ \emph {et~al.}(1969)\citenamefont {Arima},
  \citenamefont {Harvey},\ and\ \citenamefont {Shimizu}}]{arimapsu3}%
  \BibitemOpen
  \bibfield  {author} {\bibinfo {author} {\bibfnamefont {A.}~\bibnamefont
  {Arima}}, \bibinfo {author} {\bibfnamefont {M.}~\bibnamefont {Harvey}},\ and\
  \bibinfo {author} {\bibfnamefont {K.}~\bibnamefont {Shimizu}},\ }\bibfield
  {title} {\bibinfo {title} {\textrm{Pseudo $LS$ coupling and pseudo $SU_3$
  coupling schemes}},\ }\href@noop {} {\bibfield  {journal} {\bibinfo
  {journal} {Phys. Lett. B}\ }\textbf {\bibinfo {volume} {30}},\ \bibinfo
  {pages} {517} (\bibinfo {year} {1969})}\BibitemShut {NoStop}%
\bibitem [{\citenamefont {Hecht}\ and\ \citenamefont
  {Adler}(1969)}]{hecht1969}%
  \BibitemOpen
  \bibfield  {author} {\bibinfo {author} {\bibfnamefont {K.~T.}\ \bibnamefont
  {Hecht}}\ and\ \bibinfo {author} {\bibfnamefont {A.}~\bibnamefont {Adler}},\
  }\bibfield  {title} {\bibinfo {title} {\textrm{Generalized seniority for
  favored $J\neq 0$ pairs in mixed configurations}},\ }\href@noop {} {\bibfield
   {journal} {\bibinfo  {journal} {Nucl. Phys. A}\ }\textbf {\bibinfo {volume}
  {137}},\ \bibinfo {pages} {129} (\bibinfo {year} {1969})}\BibitemShut
  {NoStop}%
\bibitem [{\citenamefont {Ekstr\"om}\ \emph {et~al.}(2009)\citenamefont
  {Ekstr\"om}, \citenamefont {Cederk\"all}, \citenamefont {DiJulio},
  \citenamefont {Fahlander},\ and\ \citenamefont {Hjorth-Jensen}}]{Cd100-104}%
  \BibitemOpen
  \bibfield  {author} {\bibinfo {author} {\bibfnamefont {A.}~\bibnamefont
  {Ekstr\"om}}, \bibinfo {author} {\bibfnamefont {J.}~\bibnamefont
  {Cederk\"all}}, \bibinfo {author} {\bibfnamefont {D.~D.}\ \bibnamefont
  {DiJulio}}, \bibinfo {author} {\bibfnamefont {C.}~\bibnamefont {Fahlander}},\
  and\ \bibinfo {author} {\bibfnamefont {M.}~\bibnamefont {Hjorth-Jensen}},\
  }\bibfield  {title} {\bibinfo {title} {\textrm{Electric quadrupole moments of
  the ${2}_{1}^{+}$ states in $^{100,102,104}$Cd}},\ }\href@noop {} {\bibfield
  {journal} {\bibinfo  {journal} {Phys. Rev. C}\ }\textbf {\bibinfo {volume}
  {80}},\ \bibinfo {pages} {054302} (\bibinfo {year} {2009})}\BibitemShut
  {NoStop}%
\bibitem [{\citenamefont {Boelaert}\ \emph
  {et~al.}(2007{\natexlab{a}})\citenamefont {Boelaert}, \citenamefont {Dewald},
  \citenamefont {Fransen}, \citenamefont {Jolie}, \citenamefont {Linnemann},
  \citenamefont {Melon}, \citenamefont {M\"oller}, \citenamefont {Smirnova},\
  and\ \citenamefont {Heyde}}]{Cd102-104}%
  \BibitemOpen
  \bibfield  {author} {\bibinfo {author} {\bibfnamefont {N.}~\bibnamefont
  {Boelaert}}, \bibinfo {author} {\bibfnamefont {A.}~\bibnamefont {Dewald}},
  \bibinfo {author} {\bibfnamefont {C.}~\bibnamefont {Fransen}}, \bibinfo
  {author} {\bibfnamefont {J.}~\bibnamefont {Jolie}}, \bibinfo {author}
  {\bibfnamefont {A.}~\bibnamefont {Linnemann}}, \bibinfo {author}
  {\bibfnamefont {B.}~\bibnamefont {Melon}}, \bibinfo {author} {\bibfnamefont
  {O.}~\bibnamefont {M\"oller}}, \bibinfo {author} {\bibfnamefont
  {N.}~\bibnamefont {Smirnova}},\ and\ \bibinfo {author} {\bibfnamefont
  {K.}~\bibnamefont {Heyde}},\ }\bibfield  {title} {\bibinfo {title}
  {\textrm{Low-spin electromagnetic transition probabilities in
  $^{102,104}$Cd}},\ }\href@noop {} {\bibfield  {journal} {\bibinfo  {journal}
  {Phys. Rev. C}\ }\textbf {\bibinfo {volume} {75}},\ \bibinfo {pages} {054311}
  (\bibinfo {year} {2007}{\natexlab{a}})},\ \bibinfo {note} {~erratum
  \textit{ibid.} {\bf 77}, 019901 (2008)}\BibitemShut {NoStop}%
\bibitem [{\citenamefont {Bogner}\ \emph {et~al.}(2003)\citenamefont {Bogner},
  \citenamefont {Kuo},\ and\ \citenamefont {Schwenk}}]{vlk}%
  \BibitemOpen
  \bibfield  {author} {\bibinfo {author} {\bibfnamefont {S.}~\bibnamefont
  {Bogner}}, \bibinfo {author} {\bibfnamefont {T.~T.~S.}\ \bibnamefont {Kuo}},\
  and\ \bibinfo {author} {\bibfnamefont {A.}~\bibnamefont {Schwenk}},\
  }\bibfield  {title} {\bibinfo {title} {Model-independent low momentum nucleon
  interaction from phase shift equivalence},\ }\href@noop {} {\bibfield
  {journal} {\bibinfo  {journal} {Phys. Rep.}\ }\textbf {\bibinfo {volume}
  {386}},\ \bibinfo {pages} {1} (\bibinfo {year} {2003})}\BibitemShut {NoStop}%
\bibitem [{\citenamefont {Entem}\ and\ \citenamefont
  {Machleidt}(2002)}]{N3LOa}%
  \BibitemOpen
  \bibfield  {author} {\bibinfo {author} {\bibfnamefont {D.~R.}\ \bibnamefont
  {Entem}}\ and\ \bibinfo {author} {\bibfnamefont {R.}~\bibnamefont
  {Machleidt}},\ }\bibfield  {title} {\bibinfo {title} {Accurate
  nucleon-nucleon potential based upon chiral perturbation theory},\
  }\href@noop {} {\bibfield  {journal} {\bibinfo  {journal} {Phys. Lett. B}\
  }\textbf {\bibinfo {volume} {524}},\ \bibinfo {pages} {93} (\bibinfo {year}
  {2002})}\BibitemShut {NoStop}%
\bibitem [{\citenamefont {Duflo}\ and\ \citenamefont {Zuker}(1999)}]{gemo}%
  \BibitemOpen
  \bibfield  {author} {\bibinfo {author} {\bibfnamefont {J.}~\bibnamefont
  {Duflo}}\ and\ \bibinfo {author} {\bibfnamefont {A.~P.}\ \bibnamefont
  {Zuker}},\ }\bibfield  {title} {\bibinfo {title} {\textrm{The nuclear
  monopole Hamiltonian}},\ }\href@noop {} {\bibfield  {journal} {\bibinfo
  {journal} {Phys. Rev. C}\ }\textbf {\bibinfo {volume} {59}},\ \bibinfo
  {pages} {R2347} (\bibinfo {year} {1999})},\ \bibinfo {note} {program
  gemosp9.f in the archive Duflo-Zuker-program.zip from
  https://www-nds.iaea.org/amdc/}\BibitemShut {NoStop}%
\bibitem [{\citenamefont {Faestermann}\ \emph {et~al.}(2013)\citenamefont
  {Faestermann}, \citenamefont {G\'orska},\ and\ \citenamefont
  {Grawe}}]{Sn100}%
  \BibitemOpen
  \bibfield  {author} {\bibinfo {author} {\bibfnamefont {T.}~\bibnamefont
  {Faestermann}}, \bibinfo {author} {\bibfnamefont {M.}~\bibnamefont
  {G\'orska}},\ and\ \bibinfo {author} {\bibfnamefont {H.}~\bibnamefont
  {Grawe}},\ }\bibfield  {title} {\bibinfo {title} {\textrm{The structure of
  $^{100}$Sn and neighbouring nuclei}},\ }\href@noop {} {\bibfield  {journal}
  {\bibinfo  {journal} {Prog. Part. Nucl. Phys.}\ }\textbf {\bibinfo {volume}
  {69}},\ \bibinfo {pages} {85} (\bibinfo {year} {2013})}\BibitemShut {NoStop}%
\bibitem [{\citenamefont {Dufour}\ and\ \citenamefont {Zuker}(1996)}]{mdz}%
  \BibitemOpen
  \bibfield  {author} {\bibinfo {author} {\bibfnamefont {M.}~\bibnamefont
  {Dufour}}\ and\ \bibinfo {author} {\bibfnamefont {A.~P.}\ \bibnamefont
  {Zuker}},\ }\bibfield  {title} {\bibinfo {title} {\textrm{Realistic
  collective nuclear Hamiltonian}},\ }\href@noop {} {\bibfield  {journal}
  {\bibinfo  {journal} {Phys. Rev. C}\ }\textbf {\bibinfo {volume} {54}},\
  \bibinfo {pages} {1641} (\bibinfo {year} {1996})}\BibitemShut {NoStop}%
\bibitem [{\citenamefont {Crawford}\ \emph {et~al.}()\citenamefont {Crawford},
  \citenamefont {Clark}, \citenamefont {Fallon},\ and\ \citenamefont
  {Macchiavelli}}]{BE2FeCr}%
  \BibitemOpen
  \bibfield  {author} {\bibinfo {author} {\bibfnamefont {H.~L.}\ \bibnamefont
  {Crawford}}, \bibinfo {author} {\bibfnamefont {R.~M.}\ \bibnamefont {Clark}},
  \bibinfo {author} {\bibfnamefont {P.}~\bibnamefont {Fallon}},\ and\ \bibinfo
  {author} {\bibfnamefont {A.~O.}\ \bibnamefont {Macchiavelli}},\ }\bibfield
  {title} {\bibinfo {title} {\textrm{Quadrupole collectivity in neutron-rich Fe
  and Cr isotopes}},\ }\href@noop {} {\bibfield  {journal} {\bibinfo  {journal}
  {Phys. Rev. Lett.}\ }\textbf {\bibinfo {volume} {110}},\ \bibinfo {pages}
  {242701}}\BibitemShut {NoStop}%
\bibitem [{\citenamefont {Blazhev}\ \emph {et~al.}(2004)\citenamefont
  {Blazhev}, \citenamefont {G\'orska}, \citenamefont {Grawe}, \citenamefont
  {Nyberg}, \citenamefont {Palacz}, \citenamefont {Caurier}, \citenamefont
  {Dorvaux}, \citenamefont {Gadea},\ and\ \citenamefont
  {Nowacki}}]{PhysRevC.69.064304}%
  \BibitemOpen
  \bibfield  {author} {\bibinfo {author} {\bibfnamefont {A.}~\bibnamefont
  {Blazhev}}, \bibinfo {author} {\bibfnamefont {M.}~\bibnamefont {G\'orska}},
  \bibinfo {author} {\bibfnamefont {H.}~\bibnamefont {Grawe}}, \bibinfo
  {author} {\bibfnamefont {J.}~\bibnamefont {Nyberg}}, \bibinfo {author}
  {\bibfnamefont {M.}~\bibnamefont {Palacz}}, \bibinfo {author} {\bibfnamefont
  {E.}~\bibnamefont {Caurier}}, \bibinfo {author} {\bibfnamefont
  {O.}~\bibnamefont {Dorvaux}}, \bibinfo {author} {\bibfnamefont
  {A.}~\bibnamefont {Gadea}},\ and\ \bibinfo {author} {\bibfnamefont
  {F.}~\bibnamefont {Nowacki}},\ }\bibfield  {title} {\bibinfo {title}
  {\textrm{Observation of a core-excited $E4$ isomer in $^{98}$Cd}},\
  }\href@noop {} {\bibfield  {journal} {\bibinfo  {journal} {Phys. Rev. C}\
  }\textbf {\bibinfo {volume} {69}},\ \bibinfo {pages} {064304} (\bibinfo
  {year} {2004})}\BibitemShut {NoStop}%
\bibitem [{nud()}]{nudat2}%
  \BibitemOpen
  \href@noop {} {}\bibinfo {note}
  {\textrm{https://www.nndc.bnl.gov/nudat2}}\BibitemShut {NoStop}%
\bibitem [{\citenamefont {Schmidt}\ \emph {et~al.}(2017)\citenamefont
  {Schmidt}, , \citenamefont {Heyde}, \citenamefont {Blahzev},\ and\
  \citenamefont {Jolie}}]{schmidt2017}%
  \BibitemOpen
  \bibfield  {author} {\bibinfo {author} {\bibfnamefont {T.}~\bibnamefont
  {Schmidt}}, , \bibinfo {author} {\bibfnamefont {K.~L.~G.}\ \bibnamefont
  {Heyde}}, \bibinfo {author} {\bibfnamefont {A.}~\bibnamefont {Blahzev}},\
  and\ \bibinfo {author} {\bibfnamefont {J.}~\bibnamefont {Jolie}},\ }\bibfield
   {title} {\bibinfo {title} {Shell-model-based deformation analysis of light
  cadmium isotopes},\ }\href@noop {} {\bibfield  {journal} {\bibinfo  {journal}
  {Phys. Rev. C}\ }\textbf {\bibinfo {volume} {96}},\ \bibinfo {pages} {014302}
  (\bibinfo {year} {2017})}\BibitemShut {NoStop}%
\bibitem [{\citenamefont {Boelaert}\ \emph
  {et~al.}(2007{\natexlab{b}})\citenamefont {Boelaert}, \citenamefont
  {Smirnova}, \citenamefont {Heyde},\ and\ \citenamefont {Jolie}}]{Cdth}%
  \BibitemOpen
  \bibfield  {author} {\bibinfo {author} {\bibfnamefont {N.}~\bibnamefont
  {Boelaert}}, \bibinfo {author} {\bibfnamefont {N.}~\bibnamefont {Smirnova}},
  \bibinfo {author} {\bibfnamefont {K.}~\bibnamefont {Heyde}},\ and\ \bibinfo
  {author} {\bibfnamefont {J.}~\bibnamefont {Jolie}},\ }\bibfield  {title}
  {\bibinfo {title} {\textrm{Shell model description of the low-lying states of
  light cadmium isotopes}},\ }\href@noop {} {\bibfield  {journal} {\bibinfo
  {journal} {Phys. Rev. C}\ }\textbf {\bibinfo {volume} {75}},\ \bibinfo
  {pages} {014316} (\bibinfo {year} {2007}{\natexlab{b}})}\BibitemShut
  {NoStop}%
\bibitem [{\citenamefont {Machleidt}\ \emph {et~al.}(1996)\citenamefont
  {Machleidt}, \citenamefont {Sammarruca},\ and\ \citenamefont {Song}}]{CDB}%
  \BibitemOpen
  \bibfield  {author} {\bibinfo {author} {\bibfnamefont {R.}~\bibnamefont
  {Machleidt}}, \bibinfo {author} {\bibfnamefont {F.}~\bibnamefont
  {Sammarruca}},\ and\ \bibinfo {author} {\bibfnamefont {Y.}~\bibnamefont
  {Song}},\ }\bibfield  {title} {\bibinfo {title} {Nonlocal nature of the
  nuclear force and its impact on nuclear structure},\ }\href@noop {}
  {\bibfield  {journal} {\bibinfo  {journal} {Phys. Rev. C}\ }\textbf {\bibinfo
  {volume} {53}},\ \bibinfo {pages} {R1483} (\bibinfo {year}
  {1996})}\BibitemShut {NoStop}%
\bibitem [{\citenamefont {Smirnova}()}]{nadya}%
  \BibitemOpen
  \bibfield  {author} {\bibinfo {author} {\bibfnamefont {N.}~\bibnamefont
  {Smirnova}},\ }\href@noop {} {\bibinfo {title} {The v3sb interaction}},\
  \bibinfo {note} {~private communication (2017)}\BibitemShut {NoStop}%
\bibitem [{\citenamefont {Siciliano}\ \emph {et~al.}(2020)\citenamefont
  {Siciliano}, \citenamefont {Valiente-Dobón}, \citenamefont {Goasduff},
  \citenamefont {Nowacki}, \citenamefont {Zuker} \emph
  {et~al.}}]{siciliano2020}%
  \BibitemOpen
  \bibfield  {author} {\bibinfo {author} {\bibfnamefont {M.}~\bibnamefont
  {Siciliano}}, \bibinfo {author} {\bibfnamefont {J.~J.}\ \bibnamefont
  {Valiente-Dobón}}, \bibinfo {author} {\bibfnamefont {A.}~\bibnamefont
  {Goasduff}}, \bibinfo {author} {\bibfnamefont {F.}~\bibnamefont {Nowacki}},
  \bibinfo {author} {\bibfnamefont {A.~P.}\ \bibnamefont {Zuker}}, \emph
  {et~al.},\ }\bibfield  {title} {\bibinfo {title} {\textrm{Pairing-quadrupole
  interplay in the neutron-deficient tin nuclei: First lifetime measurements of
  low-lying states in $^{106,108}$Sn}},\ }\href@noop {} {\bibfield  {journal}
  {\bibinfo  {journal} {Phys. Lett. B}\ }\textbf {\bibinfo {volume} {806}},\
  \bibinfo {pages} {135474} (\bibinfo {year} {2020})}\BibitemShut {NoStop}%
\bibitem [{\citenamefont {de~Angelis}\ \emph {et~al.}(2002)\citenamefont
  {de~Angelis}, \citenamefont {Gadea}, \citenamefont {Farnea}, \citenamefont
  {Isocrate}, \citenamefont {Petkov}, \citenamefont {Marginean}, \citenamefont
  {Napoli}, \citenamefont {Dewald} \emph {et~al.}}]{deangelis2002coherent}%
  \BibitemOpen
  \bibfield  {author} {\bibinfo {author} {\bibfnamefont {G.}~\bibnamefont
  {de~Angelis}}, \bibinfo {author} {\bibfnamefont {A.}~\bibnamefont {Gadea}},
  \bibinfo {author} {\bibfnamefont {E.}~\bibnamefont {Farnea}}, \bibinfo
  {author} {\bibfnamefont {R.}~\bibnamefont {Isocrate}}, \bibinfo {author}
  {\bibfnamefont {P.}~\bibnamefont {Petkov}}, \bibinfo {author} {\bibfnamefont
  {N.}~\bibnamefont {Marginean}}, \bibinfo {author} {\bibfnamefont
  {D.}~\bibnamefont {Napoli}}, \bibinfo {author} {\bibfnamefont
  {A.}~\bibnamefont {Dewald}}, \emph {et~al.},\ }\bibfield  {title} {\bibinfo
  {title} {\textrm{Coherent proton–neutron contribution to octupole
  correlations in the neutron-deficient $^{114}$Xe nucleus}},\ }\href@noop {}
  {\bibfield  {journal} {\bibinfo  {journal} {Phys. Lett. B}\ }\textbf
  {\bibinfo {volume} {535}},\ \bibinfo {pages} {93} (\bibinfo {year}
  {2002})}\BibitemShut {NoStop}%
\bibitem [{\citenamefont {M{\"o}ller}\ \emph {et~al.}(2005)\citenamefont
  {M{\"o}ller}, \citenamefont {Warr}, \citenamefont {Jolie}, \citenamefont
  {Dewald}, \citenamefont {Fitzler}, \citenamefont {Linnemann}, \citenamefont
  {Zell}, \citenamefont {Garrett},\ and\ \citenamefont {Yates}}]{moller2005e2}%
  \BibitemOpen
  \bibfield  {author} {\bibinfo {author} {\bibfnamefont {O.}~\bibnamefont
  {M{\"o}ller}}, \bibinfo {author} {\bibfnamefont {N.}~\bibnamefont {Warr}},
  \bibinfo {author} {\bibfnamefont {J.}~\bibnamefont {Jolie}}, \bibinfo
  {author} {\bibfnamefont {A.}~\bibnamefont {Dewald}}, \bibinfo {author}
  {\bibfnamefont {A.}~\bibnamefont {Fitzler}}, \bibinfo {author} {\bibfnamefont
  {A.}~\bibnamefont {Linnemann}}, \bibinfo {author} {\bibfnamefont {K.~O.}\
  \bibnamefont {Zell}}, \bibinfo {author} {\bibfnamefont {P.~E.}\ \bibnamefont
  {Garrett}},\ and\ \bibinfo {author} {\bibfnamefont {S.~W.}\ \bibnamefont
  {Yates}},\ }\bibfield  {title} {\bibinfo {title} {\textrm{E2 transition
  probabilities in $^{114}$Te: A conundrum}},\ }\href@noop {} {\bibfield
  {journal} {\bibinfo  {journal} {Phys. Rev. C}\ }\textbf {\bibinfo {volume}
  {71}},\ \bibinfo {pages} {064324} (\bibinfo {year} {2005})}\BibitemShut
  {NoStop}%
\bibitem [{\citenamefont {Cederwall}\ \emph {et~al.}(2018)\citenamefont
  {Cederwall}, \citenamefont {Doncel}, \citenamefont {Aktas}, \citenamefont
  {Ertoprak}, \citenamefont {Liotta}, \citenamefont {Qi}, \citenamefont
  {Grahn}, \citenamefont {Cullen}, \citenamefont {Nara~Singh}, \citenamefont
  {Hodge}, \citenamefont {Giles},\ and\ \citenamefont
  {Stolze}}]{cederwall2018}%
  \BibitemOpen
  \bibfield  {author} {\bibinfo {author} {\bibfnamefont {B.}~\bibnamefont
  {Cederwall}}, \bibinfo {author} {\bibfnamefont {M.}~\bibnamefont {Doncel}},
  \bibinfo {author} {\bibfnamefont {O.}~\bibnamefont {Aktas}}, \bibinfo
  {author} {\bibfnamefont {A.}~\bibnamefont {Ertoprak}}, \bibinfo {author}
  {\bibfnamefont {R.}~\bibnamefont {Liotta}}, \bibinfo {author} {\bibfnamefont
  {C.}~\bibnamefont {Qi}}, \bibinfo {author} {\bibfnamefont {T.}~\bibnamefont
  {Grahn}}, \bibinfo {author} {\bibfnamefont {D.~M.}\ \bibnamefont {Cullen}},
  \bibinfo {author} {\bibfnamefont {B.~S.}\ \bibnamefont {Nara~Singh}},
  \bibinfo {author} {\bibfnamefont {D.}~\bibnamefont {Hodge}}, \bibinfo
  {author} {\bibfnamefont {M.}~\bibnamefont {Giles}},\ and\ \bibinfo {author}
  {\bibfnamefont {S.}~\bibnamefont {Stolze}},\ }\bibfield  {title} {\bibinfo
  {title} {\textrm{Lifetime measurements of excited states in $^{172}$Pt and
  the variation of quadrupole transition strength with angular momentum}},\
  }\href@noop {} {\bibfield  {journal} {\bibinfo  {journal} {Phys. Rev. Lett.}\
  }\textbf {\bibinfo {volume} {121}},\ \bibinfo {pages} {022502} (\bibinfo
  {year} {2018})}\BibitemShut {NoStop}%
\bibitem [{\citenamefont {B\"ack}\ \emph {et~al.}(2013)\citenamefont {B\"ack},
  \citenamefont {Qi}, \citenamefont {Cederwall}, \citenamefont {Liotta},
  \citenamefont {Ghazi~Moradi}, \citenamefont {Johnson}, \citenamefont {Wyss},\
  and\ \citenamefont {Wadsworth}}]{BE2Sn100}%
  \BibitemOpen
  \bibfield  {author} {\bibinfo {author} {\bibfnamefont {T.}~\bibnamefont
  {B\"ack}}, \bibinfo {author} {\bibfnamefont {C.}~\bibnamefont {Qi}}, \bibinfo
  {author} {\bibfnamefont {B.}~\bibnamefont {Cederwall}}, \bibinfo {author}
  {\bibfnamefont {R.}~\bibnamefont {Liotta}}, \bibinfo {author} {\bibfnamefont
  {F.}~\bibnamefont {Ghazi~Moradi}}, \bibinfo {author} {\bibfnamefont
  {A.}~\bibnamefont {Johnson}}, \bibinfo {author} {\bibfnamefont
  {R.}~\bibnamefont {Wyss}},\ and\ \bibinfo {author} {\bibfnamefont
  {R.}~\bibnamefont {Wadsworth}},\ }\bibfield  {title} {\bibinfo {title}
  {\textrm{Transition probabilities near $^{100}$Sn and the stability of the
  $N,Z=50$ shell closure}},\ }\href@noop {} {\bibfield  {journal} {\bibinfo
  {journal} {Phys. Rev. C}\ }\textbf {\bibinfo {volume} {87}},\ \bibinfo
  {pages} {031306(R)} (\bibinfo {year} {2013})}\BibitemShut {NoStop}%
\bibitem [{\citenamefont {Togashi}\ \emph {et~al.}(2018)\citenamefont
  {Togashi}, \citenamefont {Tsunoda}, \citenamefont {Otsuka}, \citenamefont
  {Shimizu},\ and\ \citenamefont {Honma}}]{PhysRevLett.121.062501}%
  \BibitemOpen
  \bibfield  {author} {\bibinfo {author} {\bibfnamefont {T.}~\bibnamefont
  {Togashi}}, \bibinfo {author} {\bibfnamefont {Y.}~\bibnamefont {Tsunoda}},
  \bibinfo {author} {\bibfnamefont {T.}~\bibnamefont {Otsuka}}, \bibinfo
  {author} {\bibfnamefont {N.}~\bibnamefont {Shimizu}},\ and\ \bibinfo {author}
  {\bibfnamefont {M.}~\bibnamefont {Honma}},\ }\bibfield  {title} {\bibinfo
  {title} {Novel shape evolution in \textrm{Sn} isotopes from magic numbers 50
  to 82},\ }\href@noop {} {\bibfield  {journal} {\bibinfo  {journal} {Phys.
  Rev. Lett.}\ }\textbf {\bibinfo {volume} {121}},\ \bibinfo {pages} {062501}
  (\bibinfo {year} {2018})}\BibitemShut {NoStop}%
\bibitem [{\citenamefont {Hjorth-Jensen}\ \emph {et~al.}(1995)\citenamefont
  {Hjorth-Jensen}, \citenamefont {Kuo},\ and\ \citenamefont
  {Osnes}}]{Hjorth-Jensen.Kuo.Osnes:1995}%
  \BibitemOpen
  \bibfield  {author} {\bibinfo {author} {\bibfnamefont {M.}~\bibnamefont
  {Hjorth-Jensen}}, \bibinfo {author} {\bibfnamefont {T.~T.~S.}\ \bibnamefont
  {Kuo}},\ and\ \bibinfo {author} {\bibfnamefont {E.}~\bibnamefont {Osnes}},\
  }\bibfield  {title} {\bibinfo {title} {Realistic effective interactions for
  nuclear systems},\ }\href@noop {} {\bibfield  {journal} {\bibinfo  {journal}
  {Phys. Rep.}\ }\textbf {\bibinfo {volume} {261}},\ \bibinfo {pages} {126}
  (\bibinfo {year} {1995})}\BibitemShut {NoStop}%
\bibitem [{\citenamefont {Beck}\ \emph {et~al.}(1987)\citenamefont {Beck},
  \citenamefont {Eder}, \citenamefont {Hagn},\ and\ \citenamefont
  {Zech}}]{neutron-gl}%
  \BibitemOpen
  \bibfield  {author} {\bibinfo {author} {\bibfnamefont {R.}~\bibnamefont
  {Beck}}, \bibinfo {author} {\bibfnamefont {R.}~\bibnamefont {Eder}}, \bibinfo
  {author} {\bibfnamefont {E.}~\bibnamefont {Hagn}},\ and\ \bibinfo {author}
  {\bibfnamefont {E.}~\bibnamefont {Zech}},\ }\bibfield  {title} {\bibinfo
  {title} {\textrm{Measurement of the anomalous neutron orbital $g$ factor in
  $^{190m}$Os}},\ }\href@noop {} {\bibfield  {journal} {\bibinfo  {journal}
  {Phys. Rev. Lett.}\ }\textbf {\bibinfo {volume} {59}},\ \bibinfo {pages}
  {2923} (\bibinfo {year} {1987})}\BibitemShut {NoStop}%
\bibitem [{\citenamefont {Allmond}\ \emph {et~al.}(2015)\citenamefont
  {Allmond}, \citenamefont {Stuchbery}, \citenamefont {Galindo-Uribarri},
  \citenamefont {Padilla-Rodal},\ and\ \citenamefont
  {Radford}}]{PhysRevC.92.041303}%
  \BibitemOpen
  \bibfield  {author} {\bibinfo {author} {\bibfnamefont {J.~M.}\ \bibnamefont
  {Allmond}}, \bibinfo {author} {\bibfnamefont {A.~E.}\ \bibnamefont
  {Stuchbery}}, \bibinfo {author} {\bibfnamefont {A.}~\bibnamefont
  {Galindo-Uribarri}}, \bibinfo {author} {\bibfnamefont {E.}~\bibnamefont
  {Padilla-Rodal}},\ and\ \bibinfo {author} {\bibfnamefont {D.}~\bibnamefont
  {Radford}},\ }\bibfield  {title} {\bibinfo {title} {Investigation into the
  semimagic nature of the tin isotopes through electromagnetic moments},\
  }\href@noop {} {\bibfield  {journal} {\bibinfo  {journal} {Phys. Rev. C}\
  }\textbf {\bibinfo {volume} {92}},\ \bibinfo {pages} {041303(R)} (\bibinfo
  {year} {2015})}\BibitemShut {NoStop}%
\bibitem [{\citenamefont {Cavanagh}\ \emph {et~al.}(1970)\citenamefont
  {Cavanagh}, \citenamefont {Coleman}, \citenamefont {Hardacre}, \citenamefont
  {Gard},\ and\ \citenamefont {Turner}}]{CAVANAGH197097}%
  \BibitemOpen
  \bibfield  {author} {\bibinfo {author} {\bibfnamefont {P.~E.}\ \bibnamefont
  {Cavanagh}}, \bibinfo {author} {\bibfnamefont {C.~F.}\ \bibnamefont
  {Coleman}}, \bibinfo {author} {\bibfnamefont {A.~G.}\ \bibnamefont
  {Hardacre}}, \bibinfo {author} {\bibfnamefont {G.~A.}\ \bibnamefont {Gard}},\
  and\ \bibinfo {author} {\bibfnamefont {J.~F.}\ \bibnamefont {Turner}},\
  }\bibfield  {title} {\bibinfo {title} {A study of the nuclear structure of
  the odd tin isotopes by means of the (p, d) reaction},\ }\href@noop {}
  {\bibfield  {journal} {\bibinfo  {journal} {Nucl. Phys. A}\ }\textbf
  {\bibinfo {volume} {141}},\ \bibinfo {pages} {97} (\bibinfo {year}
  {1970})}\BibitemShut {NoStop}%
\bibitem [{\citenamefont {Nowacki}(2019)}]{fred}%
  \BibitemOpen
  \bibfield  {author} {\bibinfo {author} {\bibfnamefont {F.}~\bibnamefont
  {Nowacki}},\ }\href@noop {} {\bibinfo {title} {Sn calculations in $sdgh$
  neutron spaces}},\ \bibinfo {howpublished} {private communication} (\bibinfo
  {year} {2019})\BibitemShut {NoStop}%
\bibitem [{\citenamefont {Stone}(2016)}]{STONE2016}%
  \BibitemOpen
  \bibfield  {author} {\bibinfo {author} {\bibfnamefont {N.}~\bibnamefont
  {Stone}},\ }\bibfield  {title} {\bibinfo {title} {Table of nuclear electric
  quadrupole moments},\ }\href@noop {} {\bibfield  {journal} {\bibinfo
  {journal} {At. Data Nucl. Data Tables}\ }\textbf {\bibinfo {volume}
  {111-112}},\ \bibinfo {pages} {1} (\bibinfo {year} {2016})}\BibitemShut
  {NoStop}%
\bibitem [{\citenamefont {Tamura}\ and\ \citenamefont
  {Udagawa}(1966)}]{tamura1966}%
  \BibitemOpen
  \bibfield  {author} {\bibinfo {author} {\bibfnamefont {T.}~\bibnamefont
  {Tamura}}\ and\ \bibinfo {author} {\bibfnamefont {T.}~\bibnamefont
  {Udagawa}},\ }\bibfield  {title} {\bibinfo {title} {\textrm{Static quadrupole
  moment of the first $2^+$ state of vibrational nuclei}},\ }\href@noop {}
  {\bibfield  {journal} {\bibinfo  {journal} {Phys. Rev.}\ }\textbf {\bibinfo
  {volume} {150}},\ \bibinfo {pages} {783} (\bibinfo {year}
  {1966})}\BibitemShut {NoStop}%
\bibitem [{\citenamefont {Garrett}\ and\ \citenamefont
  {Wood}(2010)}]{garrett2010}%
  \BibitemOpen
  \bibfield  {author} {\bibinfo {author} {\bibfnamefont {P.~E.}\ \bibnamefont
  {Garrett}}\ and\ \bibinfo {author} {\bibfnamefont {J.~L.}\ \bibnamefont
  {Wood}},\ }\bibfield  {title} {\bibinfo {title} {\textrm{On the robustness of
  surface vibrational modes: case studies in the Cd region}},\ }\href@noop {}
  {\bibfield  {journal} {\bibinfo  {journal} {J. Phys. G: Nucl. Part. Phys.}\
  }\textbf {\bibinfo {volume} {37}},\ \bibinfo {pages} {064028} (\bibinfo
  {year} {2010})}\BibitemShut {NoStop}%
\bibitem [{\citenamefont {Heyde}\ and\ \citenamefont {Wood}(2011)}]{heyde2011}%
  \BibitemOpen
  \bibfield  {author} {\bibinfo {author} {\bibfnamefont {K.}~\bibnamefont
  {Heyde}}\ and\ \bibinfo {author} {\bibfnamefont {J.~L.}\ \bibnamefont
  {Wood}},\ }\bibfield  {title} {\bibinfo {title} {Shape coexistence in atomic
  nuclei},\ }\href@noop {} {\bibfield  {journal} {\bibinfo  {journal} {Rev.
  Mod. Phys.}\ }\textbf {\bibinfo {volume} {83}},\ \bibinfo {pages} {1467}
  (\bibinfo {year} {2011})}\BibitemShut {NoStop}%
\bibitem [{\citenamefont {Leviatan}\ \emph {et~al.}(2018)\citenamefont
  {Leviatan}, \citenamefont {Gavrielov}, \citenamefont {Garc\'{\i}a-Ramos},\
  and\ \citenamefont {Van~Isacker}}]{leviatan2018}%
  \BibitemOpen
  \bibfield  {author} {\bibinfo {author} {\bibfnamefont {A.}~\bibnamefont
  {Leviatan}}, \bibinfo {author} {\bibfnamefont {N.}~\bibnamefont {Gavrielov}},
  \bibinfo {author} {\bibfnamefont {J.~E.}\ \bibnamefont {Garc\'{\i}a-Ramos}},\
  and\ \bibinfo {author} {\bibfnamefont {P.}~\bibnamefont {Van~Isacker}},\
  }\bibfield  {title} {\bibinfo {title} {Quadrupole phonons in the cadmium
  isotopes},\ }\href@noop {} {\bibfield  {journal} {\bibinfo  {journal} {Phys.
  Rev. C}\ }\textbf {\bibinfo {volume} {98}},\ \bibinfo {pages} {031302(R)}
  (\bibinfo {year} {2018})}\BibitemShut {NoStop}%
\bibitem [{\citenamefont {Garrett}\ \emph {et~al.}(2019)\citenamefont
  {Garrett}, \citenamefont {Rodr\'{\i}guez}, \citenamefont {A.}, \citenamefont
  {Green}, \citenamefont {Bangay}, \citenamefont {Finlay} \emph
  {et~al.}}]{garrett2019}%
  \BibitemOpen
  \bibfield  {author} {\bibinfo {author} {\bibfnamefont {P.~E.}\ \bibnamefont
  {Garrett}}, \bibinfo {author} {\bibfnamefont {T.~R.}\ \bibnamefont
  {Rodr\'{\i}guez}}, \bibinfo {author} {\bibfnamefont {D.}~\bibnamefont {A.}},
  \bibinfo {author} {\bibfnamefont {K.~L.}\ \bibnamefont {Green}}, \bibinfo
  {author} {\bibfnamefont {J.}~\bibnamefont {Bangay}}, \bibinfo {author}
  {\bibfnamefont {A.}~\bibnamefont {Finlay}}, \emph {et~al.},\ }\bibfield
  {title} {\bibinfo {title} {\textrm{Multiple shape coexistence in
  $^{110,112}$Cd}},\ }\href@noop {} {\bibfield  {journal} {\bibinfo  {journal}
  {Phys. Rev. Lett.}\ }\textbf {\bibinfo {volume} {123}},\ \bibinfo {pages}
  {142502} (\bibinfo {year} {2019})}\BibitemShut {NoStop}%
\end{thebibliography}%
\end{document}